# Revealing electron-lattice decoupling by Peltier thermometry and nanoscale thermal imaging in graphene

Saurabh Kumar Srivastav[1†], Tobias Völkl[1†], Gary Quaresima[2†], Yuri Myasoedov[1], Martin E. Huber[3], Kenji Watanabe[4], Takashi Taniguchi[5], L. S. Levitov[6], D. A. Pesin[2#], Eli Zeldov[1*]

Electrical currents in low-dimensional quantum materials can drive electrons far from equilibrium, creating stark imbalance between electron and lattice temperatures. Yet, no existing methods enable simultaneous nanoscale mapping of both temperatures at cryogenic conditions. Here, we introduce a scanning probe technique that images the local lattice temperature and extracts electron temperature at gate-defined p-n junctions in graphene. By applying an alternating electrical current and analyzing first- and second-harmonic responses, we disentangle Joule heating from the Peltier effect—the latter encoding the local electron temperature. This enables the first spatially resolved cryogenic imaging of both phenomena in graphene. Even under modest current bias, the electron temperature increases by nearly three orders of magnitude more than the lattice temperature, revealing strong electron–phonon decoupling and indicating a previously unrecognized electron cooling pathway. Our minimally invasive method is broadly applicable to van der Waals heterostructures and opens new avenues for probing energy dissipation and non-equilibrium transport in correlated and hydrodynamic electron systems.

[1]Department of Condensed Matter Physics, Weizmann Institute of Science, Rehovot 7610001, Israel.
[2]Department of Physics, University of Virginia, Charlottesville, Virginia 22904, USA.
[3]Departments of Physics and Electrical Engineering, University of Colorado Denver; Denver, Colorado 80217, USA.
[4]Research Center for Electronic and Optical Materials, National Institute for Materials Science, 1-1 Namiki, Tsukuba 305-0044, Japan.
[5]Research Center for Materials Nanoarchitectonics, National Institute for Materials Science, 1-1 Namiki, Tsukuba 305-0044, Japan
[6]Department of Physics, Massachusetts Institute of Technology, Cambridge, MA 02139, USA.
†These authors contributed equally to the work.
#dp7bx@virginia.edu
*eli.zeldov@weizmann.ac.il

Electronic transport in low-dimensional materials habitually results in deviations from thermal equilibrium. In high-mobility systems at cryogenic temperatures, electrons can absorb energy from an applied electric field faster than they can dissipate it to the lattice. This leads to elevated and spatially varying electron temperatures $T_e$, which can differ significantly from the local lattice temperature $T_l$. Such decoupling is especially pronounced in two-dimensional van der Waals materials, where weak electron–phonon coupling and long inelastic scattering lengths allow hot electron regimes to emerge even under modest excitation or current bias [1–9]. Accurate knowledge of both $T_e$ and $T_l$ is critical for understanding dissipation mechanisms in nanoscale devices, probing fundamental regimes such as strongly correlated phases in van der Waals heterostructures [10–15] or hydrodynamic electron flows [16–19], as well as for the technological development of sensing devices and heat dissipation mitigation in future nanoelectronics [20,21]. Yet, direct spatially resolved measurements of $T_e$ and $T_l$, particularly under cryogenic conditions, remain an outstanding experimental challenge.

A wide range of thermometry techniques has been developed to probe either the lattice or electron temperature, but rarely both. Methods such as scanning thermal microscopy [22–24], Raman thermometry [25], scanning near-field optical microscopy [26], and diamond nitrogen vacancy (NV) thermometry [27,28] offer nanoscale mapping of the lattice temperature. However, these approaches are intrinsically insensitive to electronic temperature and lack sensitivity at cryogenic temperatures. Superconducting quantum interference device on a tip (SOT) microscopy offers highly sensitive lattice thermal imaging at cryogenic temperatures [29–31], but does not directly measure electron temperature.

In contrast, electronic thermometry techniques such as Johnson–Nyquist thermometry (JNT) [5,8,19,32–36], shot noise thermometry (SNT) [37], Coulomb blockade thermometry (CBT) [38–40], and superconducting Josephson-junction devices [38,41–44], directly probe $T_e$ at cryogenic temperatures but typically lack spatial resolution. JNT and SNT measure $T_e$ averaged over the device scale and require precise noise calibration, while CBT and Josephson thermometers depend on complex nanofabrication at specific locations. Recently, near-field terahertz scanning noise microscopy [45] enabled nanoscale mapping of $T_e$, though its sensitivity remains limited at low temperatures.

As a result, no existing technique can provide simultaneous spatial mapping of both electron and lattice temperature profiles at the nanoscale under cryogenic conditions—despite the growing need for such capability in understanding energy dissipation, heat flow, and thermal management in low-dimensional quantum systems. Here, we present a thermometry technique that overcomes these limitations by combining nanoscale imaging of the lattice temperature with local electron temperature extraction, enabling the first direct comparison between them. Our method exploits the Peltier effect, in which electrical current through a junction between materials with different thermopower generates localized heating or cooling. Crucially, the Peltier coefficient depends directly on the local electron temperature. By applying an alternating current and measuring the first and second harmonics of the resulting lattice temperature variations, we isolate the Peltier-induced odd-harmonic signal and extract $T_e$ at electrostatically controlled p-n junctions in graphene.

We demonstrate this approach in a dual-gated graphene device, where sharp junctions generate a strong local thermoelectric response. Using a nanoscale scanning superconducting Josephson junction on a tip (JOT), we image the local lattice temperature with high spatial resolution and sensitivity. The even-harmonic response reveals Joule heating of the lattice, while the simultaneously acquired odd-harmonic Peltier signal at the junction encodes the electron temperature. This enables the first spatially resolved thermal imaging of both Joule heating and the Peltier effect in graphene at low temperatures. Remarkably, even modest current biases produce electron temperature increase up to three orders of magnitude greater than the lattice temperature increase, directly confirming strong

electron–phonon decoupling and revealing a previously unrecognized electron-lattice cooling mechanism. This minimally invasive technique requires no embedded thermometers and is broadly compatible with van der Waals materials. It opens new avenues for probing nanoscale energy dissipation, electron heat transport, and non-equilibrium phenomena in a variety of systems, including those hosting strongly correlated ground states or semiconductor-superconductor architectures.

**Operating principle of Peltier electron thermometry**

The Seebeck and Peltier are complementary thermoelectric effects that relate temperature gradients and electric fields. The Seebeck effect describes an electromotive force $E = -S\nabla T$ generated by a temperature gradient, where $S$ is the Seebeck coefficient (thermopower). The Seebeck effect underlies passive thermometry: in a thermocouple, the open-circuit voltage $\Delta V = (S_L - S_R)\Delta T$ between two dissimilar conductors L and R provides a measure of the temperature difference $\Delta T$. Conversely, the Peltier effect involves active cooling or heating of a junction when an electric current $I$ is applied. The power transferred at the junction is $\dot{Q}^j_{\text{Pelt}} = \Pi_j I$, where $\Pi_j = \Pi_L - \Pi_R$ is the Peltier coefficient. The Seebeck and Peltier coefficients are related via the Thomson relation $\Pi = ST$, which holds under local electronic equilibrium, even out of global equilibrium. Traditional thermoelectric devices are made of bulk materials where electron and lattice temperatures are assumed to be equal. In contrast, our goal is to determine the out-of-equilibrium electron temperature $T_e$ in high-mobility systems at low temperatures where weak electron-phonon coupling leads to significant deviations between $T_e$ and the lattice temperature $T_l$.

Rather than using the Peltier effect in its conventional role for active temperature control [46], we propose employing it as a thermometer to probe the local out-of-equilibrium electron temperature at a junction. This novel approach to electron thermometry takes advantage of the strong sensitivity of the Peltier coefficient to the electron temperature $T_e$. The precise form of this temperature dependence is determined by the transport regime realized in the sample. In our experiments, the electron temperatures reached at the highest applied currents remain low enough for elastic scattering to dominate transport, while electron–electron collisions are sufficiently frequent to establish a local electron temperature over length scales short compared to the device length. Under these conditions, the Peltier coefficient is described by the semiclassical Mott formula that links the measured thermoelectric response and the local electron temperature [47],

$$\Pi = \frac{\pi^2 k_B^2 T_e^2}{3e} \frac{\partial \ln \sigma}{\partial n} \nu(\varepsilon_F) \equiv \tilde{\Pi} T_e^2. \tag{1}$$

Here, $k_B$ is the Boltzmann constant, $e$ is the elementary charge, $\nu(\varepsilon_F)$ is the density of states (DOS) at the Fermi energy $\varepsilon_F$ determined by the carrier density $n$, and $\sigma$ is the doping-dependent electrical conductivity. We introduced the reduced Peltier coefficient, $\tilde{\Pi}$, a material-dependent constant, to emphasize the leading $T_e^2$ temperature dependence of the full Peltier coefficient. Small corrections of order $T_e^2/\varepsilon_F^2$, arising from the temperature dependence of the conductivity, may be present but are negligible at low temperatures.

Importantly, while all externally supplied heating power is ultimately dissipated into the lattice, it causes only a slight change in the lattice temperature $T_l$, even as the electron temperature $T_e$ varies significantly. Electrons primarily act as intermediaries in energy transport, and their temperature can vary considerably depending on the strength of electron-phonon coupling. As a result, $T_e$ cannot be reliably inferred from $T_l$ without detailed modelling of transport and relaxation processes, which is often impractical. To overcome this limitation, we work with devices incorporating p-n junctions, where, as we will see, $T_l$ can still provide access to $T_e$. By applying an *ac* current and analyzing the current dependence of $T_l$, we decompose the temperature response into even and odd harmonics.

The even harmonics reflect energy input from Joule heating and offer no direct information about $T_e$, whereas the odd harmonics arise from thermoelectric effects that primarily redistribute heat within the device. These odd harmonics in $T_l$ will be shown to encode quantitative information about the electron temperature, enabling local thermometry of the electron subsystem.

To illustrate the operating principle of this Peltier thermometry, we consider a specific geometry of a graphene device in the form of a strip of length $L\left(-\frac{L}{2} < x < \frac{L}{2}\right)$ and width $W \ll L$, with an electrostatically defined p-n junction at the center ($x = 0$). This junction separates two uniformly doped left ($L$) and right ($R$) graphene sections. Metal contacts are located at $x = \pm L/2$, allowing an *ac* current $I = JW$ to be applied through the strip, where $J$ is the current density. In this geometry, in addition to the conventional even-harmonic contributions from Joule heating, three odd-harmonic heat sources emerge from thermoelectric effects. The first one is due to the p-n junction, which acts as a 1D source delivering energy to the electrons at the rate of

$$\dot{Q}^j_{\text{Pelt}} = (\Pi_L - \Pi_R)J = (\tilde{\Pi}_L - \tilde{\Pi}_R)T_e^2 J = \tilde{\Pi}_j T_e^2 J, \tag{2}$$

per unit width of the sample. Additional 1D energy sources arise at the left and right metal-graphene interfaces,

$$\dot{Q}^L_{\text{Pelt}} = (\Pi_{Au} - \Pi_L)J = (\tilde{\Pi}_{Au} - \tilde{\Pi}_L)T_0^2 J, \qquad \dot{Q}^R_{\text{Pelt}} = (\Pi_R - \Pi_{Au})J = (\tilde{\Pi}_R - \tilde{\Pi}_{Au})T_0^2 J, \tag{3}$$

where $T_0$ is the electron temperature at the Au metal contact, which we assume to be close to the base temperature. The third source is the Thomson effect [47], stemming from the gradient of the electron temperature along the device, which acts as a 2D source in the regions away from the junction with energy delivered to the electrons at a rate of

$$\dot{q}_{\text{Th}}(x) = -JT_e \nabla \frac{\Pi}{T_e} = -J\frac{\tilde{\Pi}}{2}\nabla T_e^2, \tag{4}$$

per unit area. Integrating $\dot{q}_{\text{Th}}$ over the length of the left and right sections of graphene yields the total Thomson power per unit width

$$\dot{Q}^L_{\text{Th}} = \int_{-\frac{L}{2}}^{0} \dot{q}_{\text{Th}} dx = \frac{1}{2}J\tilde{\Pi}_L(T_0^2 - T_e^2), \quad \dot{Q}^R_{\text{Th}} = \int_{0}^{\frac{L}{2}} \dot{q}_{\text{Th}} dx = \frac{1}{2}J\tilde{\Pi}_R(T_e^2 - T_0^2). \tag{5}$$

Estimating the heating effects described by Eqs. 2, 3, and 5, we conclude that the Peltier power at the Au/graphene junctions is small, while the Thomson effect contributes over the entire length of the device. Provided that the thermal relaxation length is short compared to $L$, the dominant odd-harmonic power source at the p-n junctions is the local Peltier effect described by Eq. 2. Consequently, the resulting odd-harmonic component of the lattice temperature at the junction, $T_{\text{Pelt}}$, is given by

$$T_{\text{Pelt}} = C\dot{Q}^j_{\text{Pelt}} = C\tilde{\Pi}_j T_e^2 J, \tag{6}$$

where $T_e$ is the electron temperature at the junction, and $C$ is a proportionality constant that depends on details of the thermal transport. Crucially, the value of $C$ can be experimentally determined under equilibrium conditions in the absence of external heating and at a small applied $J$. In this limit, the Peltier-induced lattice temperature is $T^0_{\text{Pelt}} = C\tilde{\Pi}_j T_0^2 J$, where $T_0$ is the equilibrium electron temperature. Hence, the out-of-equilibrium electron temperature $T_e$ at the junction in the presence of heating can be directly extracted from the ratio of the measured $T_{\text{Pelt}}$ and $T^0_{\text{Pelt}}$ as

$$T_e = T_0\sqrt{T_{\text{Pelt}}/T^0_{\text{Pelt}}} \tag{7}$$

without the need for determining the proportionality constant $C$ or the reduced Peltier coefficient $\tilde{\Pi}_j$. This result reveals the key advantage of Peltier thermometry: despite the lattice acting only as a thermal sink, its odd-harmonic response at the junction carries direct information about the local electron temperature due to the dependence of the Peltier coefficient $\Pi_j$ on $T_e$ (Eq. 1). In contrast, the even-harmonic (Joule) heating is insensitive to the electron temperature in this way. Moreover, this approach provides simultaneous imaging of the lattice temperature across the sample and measurement of the local electron temperature in the vicinity of the junction.

Lastly, we note that a distinct feature of Peltier thermometry is that it is an active measurement technique, as it requires a small probing current analogous to standard cryogenic thermometers (e.g., calibrated resistors or semiconductor diodes), which also operate under finite current bias.

**Cryogenic imaging of the lattice temperature**

For imaging the local lattice temperature, we developed a MoRe superconducting Josephson junction on a tip (junction-on-tip, JOT, Extended Data Fig. 1), operating similarly to the previously reported thermal imaging method based on superconducting quantum interference device on a tip (SQUID-on-tip) [29,48]. The JOT with apex diameter of 50 nm (Methods) and thermal sensitivity of 2.6 μK/Hz$^{1/2}$ was scanned above the sample surface at a height of $h \cong 200$ nm and at a base temperature of $T_0 = 4.3$ K, as illustrated in Fig. 1a. The thermal coupling between the JOT and the sample is established by a helium exchange gas maintained at a low pressure of $P \cong 25$ mbar. The JOT thermometry provides a very accurate measurement of the local lattice (phonon) temperature at the sample surface with nanoscale spatial resolution [29,48]. To reduce the effect of $1/f$ noise of the JOT, a sinusoidal *ac* excitation current $I$ at frequency $f = 91.73$ Hz was applied to the two-probe device and the resulting *ac* modulation in the local excess lattice temperature $T(x, y)$ relative to the base temperature $T_0$ was imaged by the scanning JOT.

**Imaging Peltier effect in graphene**

A monolayer graphene device encapsulated in hBN with three regions doped electrostatically using separate back gates was patterned into a strip of width $W = 3.4$ μm and length $L = 17.3$ μm (Methods) with two Au contacts at the left and right edges (Fig. 1c). The lattice temperature distribution $T(x, y)$ across the device has two contributions arising from thermoelectric effects (referred to as 'Peltier' below) and from Joule heating. To distinguish between these two mechanisms, we apply an *ac* current at frequency $f$ and measure the resulting local temperature oscillations at both $f$ and $2f$. Since the Peltier effect is odd in $I$, we denote the temperature signal at $f$ as $T_\text{Pelt}$, while the signal at $2f$ is denoted $T_\text{Joule}$ because Joule heating is even (quadratic) in current.

We first consider $T_\text{Joule}$ measurements. Figure 1j shows the acquired spatial map $T_\text{Joule}(x, y)$ with the entire graphene strip uniformly hole doped with the density $n = -0.39 \times 10^{12}$ cm$^{-2}$, with $I = 20$ μA (rms) driven between the left and right contacts. The highest $T_\text{Joule}$ is observed at the Au contacts consistent with the numerical simulations (Fig. 1k), as discussed below. The two small regions of enhanced $T_\text{Joule}$ in the central region of the device in Fig. 1j are the locations of bubble defects visible in Fig. 1c, where carrier scattering and electron-phonon coupling are enhanced [48]. In contrast to $T_\text{Joule}$ that is increased only by few tens of mK in the graphene strip, the electron temperature $T_e$ is greatly enhanced, as will be demonstrated below. Moreover, $T_e$ peaks at the center of the strip and gradually approaches the base temperature $T_0$ near the Au contacts in contrast to $T_\text{Joule}(x, y)$ that peaks at the metal contacts. This is a result of weak electron-phonon coupling in graphene giving rise to a large difference between $T_e$ and $T_l$ in the bulk of the sample, while within the metal contacts $T_e \approx T_l \approx T_0$ due to strong electron-phonon coupling and good thermal conductivity to the substrate.

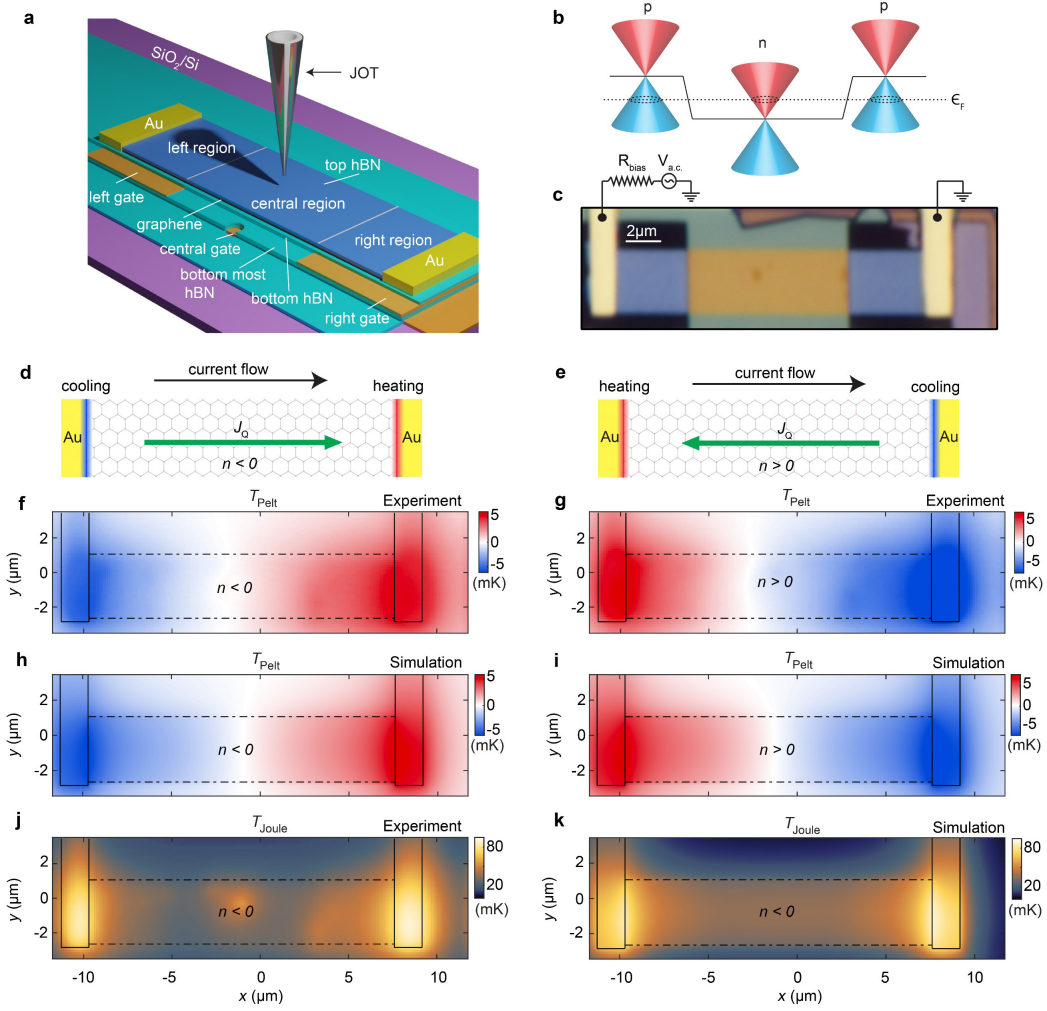

**Fig. 1. Imaging the Peltier effect at Au/graphene junctions. a,** Schematics of the hBN encapsulated graphene device with a scanning Josephson junction on a tip (JOT). Carrier doping in the center and in the left and right regions is controlled separately by local back gates. **b,** Schematics of monolayer graphene band structure in the p-n-p junction configuration in real space. The dashed line represents the position of the Fermi level in the absence of current. **c,** Optical image of the device with schematic electrical circuit added. **d,e,** Expected Peltier heating and cooling at Au/graphene junctions for uniformly p-doped (**d**) and n-doped (**e**) configurations. The green arrows depict the direction of the Peltier component of the heat current density $J_Q$. **f,** Measured map of the local Peltier lattice temperature $T_{\text{Pelt}}$ acquired by the scanning JOT at first harmonic of the applied ac current $I = 20$ μA rms at $T_0 = 4.3$ K for uniform hole doping of graphene $n = -0.39 \times 10^{12}$ cm$^{-2}$. **g,** Same as (f) for electron doping $n = 0.27 \times 10^{12}$ cm$^{-2}$. **h,i,** Numerically simulated $T_{\text{Pelt}}$ across the device (Methods) for the same current and doping values as in f,g. **j,** Measured map of local Joule lattice temperature $T_{\text{Joule}}$ acquired at the second harmonic simultaneously with (f). See Extended Data Fig. 2 for uniform electron doping configuration. **k,** Numerically simulated $T_{\text{Joule}}$ for the same conditions as in (j).

Next, we use back gates to induce a p-n-p configuration in the strip. Figure 2c shows the resulting $T_{\text{Pelt}}$ map that displays a strong Peltier effect at the p-n and n-p graphene junctions. The left junction displays large Peltier heating, in agreement with $\dot{Q}_{\text{Pelt}}^L = (\Pi_{p,\text{gr}} - \Pi_{n,\text{gr}})I > 0$, while at the right junction pronounced Peltier cooling is observed, as expected from $\dot{Q}_{\text{Pelt}}^R = (\Pi_{n,\text{gr}} - \Pi_{p,\text{gr}})I < 0$ (see a sketch in Fig. 2a). For n-p-n doping, the $T_{\text{Pelt}}$ map is of the opposite sign as shown in Figs. 2b,d. Notably, the observed $|T_{\text{Pelt}}|$ at the graphene junctions is an order of magnitude larger than at the Au/graphene junctions, as shown in Figs. 1f,g. This is a result of three effects. First, since $\tilde{\Pi}_{Au}$ is small, $\tilde{\Pi}_{p,gr} - \tilde{\Pi}_{n,gr}$ is about twice larger than $\tilde{\Pi}_{p,gr} - \tilde{\Pi}_{Au}$ giving rise to larger $\dot{Q}_{Pelt}$ at graphene junctions. Second, the thermal conductivity of Au film is much higher than that of monolayer graphene, leading to enhanced heat diffusion that suppresses the resulting $T_{Pelt}$ lattice temperature at Au/graphene junctions. Finally, the enhanced electron-phonon coupling in Au lowers $T_e$ at the Au/graphene junction, leading to a reduction of the Peltier effect as discussed below.

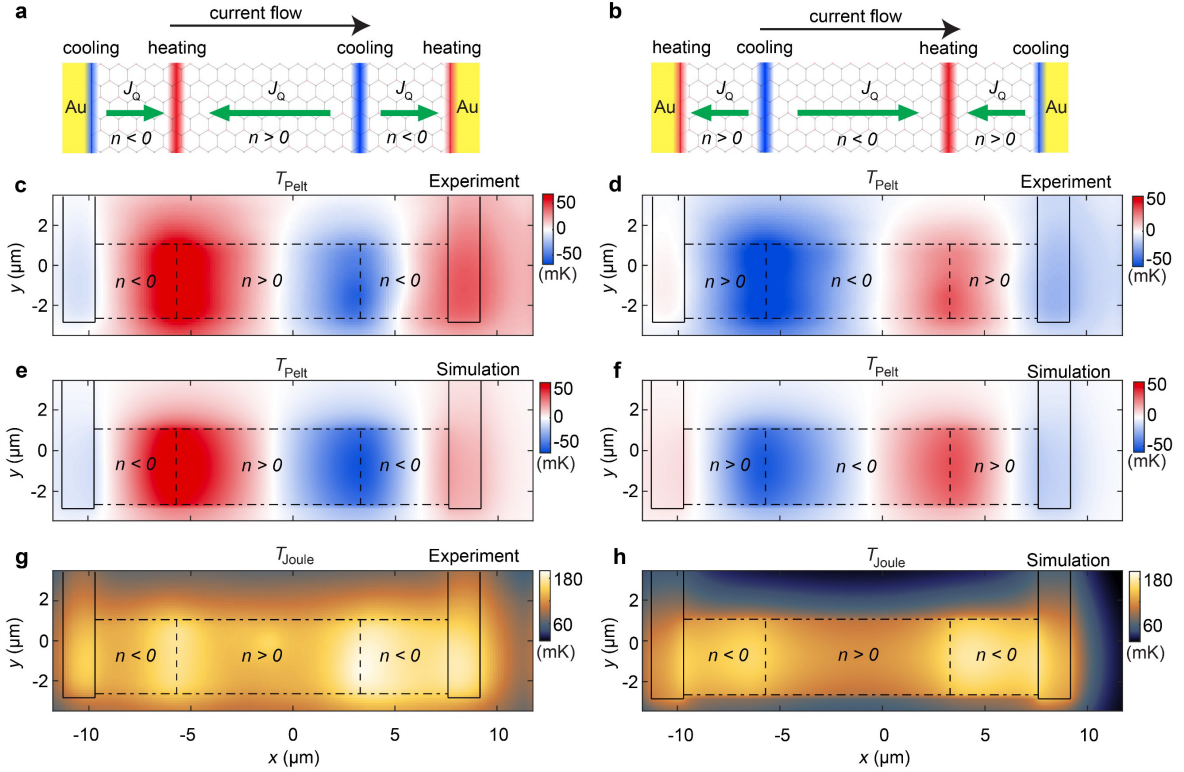

**Fig. 2. Imaging of Peltier effect at graphene p-n junctions. a,b,** Schematics of the Peltier heating (red) and cooling (blue) at p-n and n-p locations for p-n-p (**a**) and n-p-n (**b**) doping. The green arrows depict the direction of the Peltier component of the heat current density $J_Q$ in the left, center, and right graphene regions. **c,d,** Peltier lattice temperature $T_{Pelt}$ map acquired at $I = 20$ μA rms at $T_0 = 4.3$ K for the p-n-p doping of $(-0.085, 0.19, -0.085) \times 10^{12}$ cm$^{-2}$ (**c**), and n-p-n doping $(0.08, -0.29, 0.08) \times 10^{12}$ cm$^{-2}$ (**d**). **e,f,** Numerically simulated $T_{Pelt}$ (Methods) across the device for the same current and doping values as in (c,d). **g,** Joule lattice temperature $T_{Joule}$ for p-n-p doping acquired simultaneously with (c). See Extended Data Fig. 2 for n-p-n doping. **h,** Numerically simulated $T_{Joule}$ for the same conditions as in (g).

The measured $T_{Joule}$ map with p-n-p configuration, displayed in Fig. 2g, shows significantly higher overall lattice heating as compared to the uniform case in Fig. 1j. The enhanced heating is due to the lower graphene doping and hence higher resistance. Interestingly, enhanced $T_{Joule}$ is revealed near the two graphene junctions, even though there is no excess local Joule heating at the junctions. As shown below, this is a result of the local change in the electron-phonon coupling strength, consistent with the numerical simulation in Fig. 2h.

We note parenthetically that Figs. 1 and 2 constitute the first thermal imaging of the thermoelectric effects in graphene at cryogenic temperatures.

**Peltier response of a p-n junction**

To gain further insight into the Peltier effect at a single p-n junction, we performed point measurements by placing the JOT above the center of the right junction and measuring $T_{Joule}$ and $T_{\text{Pelt}}$ as a function of carrier densities and polarities in the central, $n_c$, and in the outer left and right graphene regions, while maintaining $n_l = n_r$, at a constant $I$. Figures 3a,b show a comparison between the two-probe resistance of the device $R_{2P}$ and $T_{\text{Joule}}$ as a function of $n_c$ and $n_r$. $T_{\text{Joule}}$ closely follows the $R_{2P}$ as expected for Joule dissipation. This is further visualized by comparing the various $R_{2P}$ line cuts in Figs. 3d,e with those of $T_{\text{Joule}}$ in Figs. 3f,g, showing the same qualitative behavior.

In contrast to $T_{\text{Joule}}$, $T_{\text{Pelt}}$ in Fig. 3c shows a pattern with several sign changes. While $R_{2P}$ and $T_{\text{Joule}}$ peak at the charge neutrality point (CNP) and decrease monotonically with $|n|$, $T_{\text{Pelt}}$ vanishes at CNP and shows a non-monotonic evolution around it. This response becomes more evident in the line cuts in Figs. 3h,i. The various sign changes arise from two effects. First, the Peltier coefficient $\Pi_{\text{gr}}$ changes sign between electron and hole doping. Secondly, since $\dot{Q}_{\text{Pelt}} = (\Pi_{\text{gr},c} - \Pi_{\text{gr},r})I$, the heating or cooling rate depends on the Peltier coefficient difference between the two sides of the junction ($\Pi_{\text{gr},c}$ and $\Pi_{\text{gr},r}$ in the central and right graphene regions). In the second quadrant in Fig. 3c (top-left), $\Pi_{gr,c} > 0$ and $\Pi_{gr,r} < 0$ and hence $\dot{Q}_{\text{Pelt}}$ and $T_{\text{Pelt}}$ are positive for any doping within the quadrant. Similarly, in the fourth quadrant $T_{\text{Pelt}}$ is always negative. In the first and third quadrants, in contrast, both signs of $T_{\text{Pelt}}$ can occur depending on the relative magnitudes of $\Pi_{gr,c}$ and $\Pi_{gr,r}$.

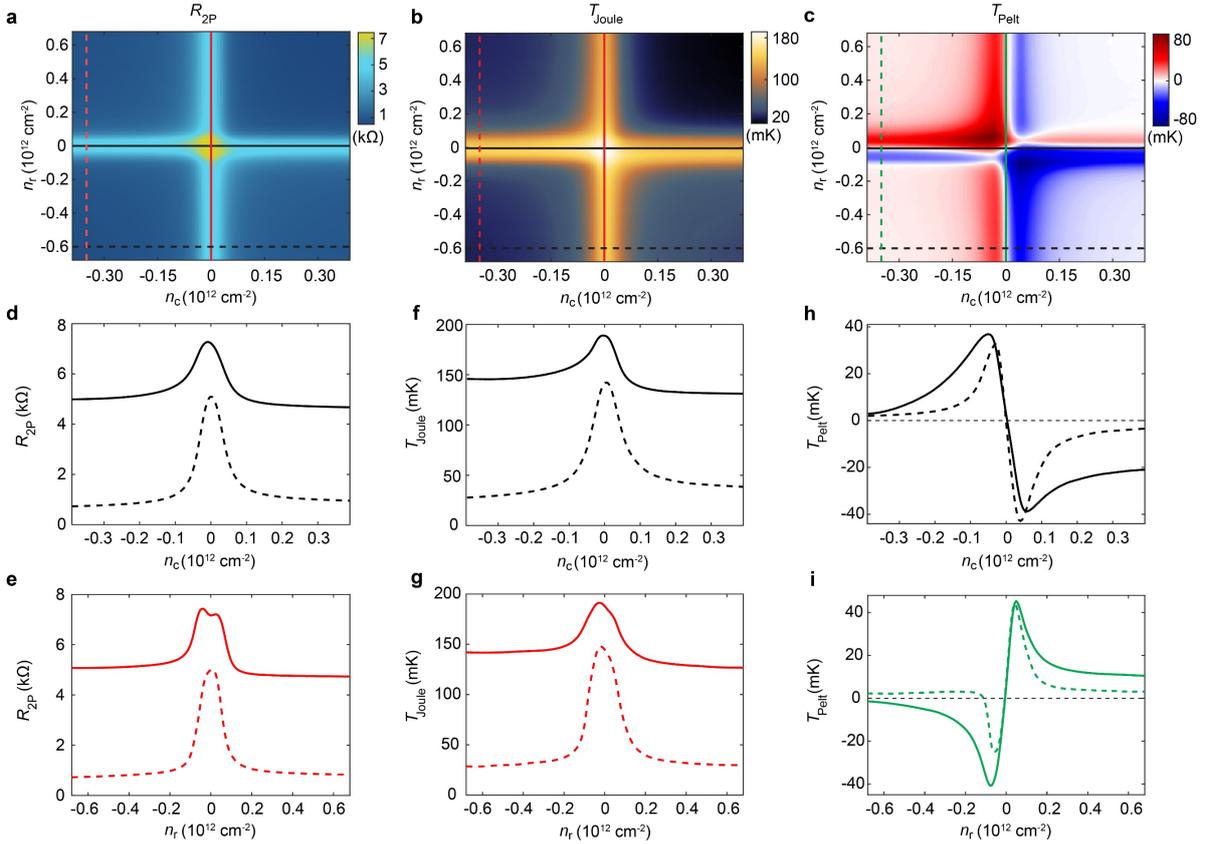

**Fig. 3. Doping dependence of Joule dissipation and Peltier effect at p-n junction. a,** Two-probe sample resistance $R_{2P}$ as a function of density in the central $n_c$ and outer $n_r = n_l$ graphene regions. **b,** $T_{\text{Joule}}$ at the right p-n junction as a function of $n_c$ and $n_r = n_l$, taken at $I = 17$ μA rms. **c,** $T_{\text{Pelt}}$ at the right p-n junction measured simultaneously with (b). **d,** Linecuts of $R_{2P}$ along the black solid and dashed horizontal lines in (a), taken at $n_r = 0$ and $-0.6 \times 10^{12}$ cm$^{-2}$. **e,** Linecuts of $R_{2P}$ along the red solid and dashed lines in (a) at $n_c = 0$ and $-0.35 \times 10^{12}$ cm$^{-2}$. **f,g,** Linecuts of $T_{\text{Joule}}$ along the black and red solid and dashed lines in (b). **h,i,** Linecuts of $T_{\text{Pelt}}$ along the black and green solid and dashed lines in (c).

Using Eq. 1 with graphene DOS $\nu(\varepsilon_F) = \frac{2}{\hbar v_F}\sqrt{\frac{|n|}{\pi}}$ and $\sigma$ estimated from the measurement of $R_{2P}$, we can evaluate the behavior of the Peltier lattice temperature at the graphene p-n junction, $T_{\text{Pelt}} \propto \dot{Q}_{\text{Pelt}} = (\Pi_{gr,c} - \Pi_{gr,r})I$. Extended Data Fig. 4 shows the calculated $\dot{Q}_{\text{Pelt}}$ based on the measured $R_{2P}$ in Fig. 3a to be in good agreement with the measured $T_{\text{Pelt}}$ in Fig. 3c, corroborating the validity of Eq. 1.

**Nonlinear Peltier effect**

Figures 4a,d show $T_{\text{Joule}}$ and $T_{\text{Pelt}}$ at the right graphene p-n junction vs. $n_c$ and $I$, while keeping the outer regions at fixed large hole doping $n_l = n_r = -0.43 \times 10^{12}$ cm$^{-2}$. The magnitudes of both $T_{\text{Joule}}$ and $T_{\text{Pelt}}$ increase with $I$. The current dependence of $T_{\text{Joule}}$ is presented in Figs. 4b,c for indicated electron and hole doping values $n_c$. As expected for Joule dissipation, $T_{\text{Joule}}$ (open circles) can be well fitted by a $I^2$ dependence (solid lines). Note that for comparable values of carrier densities, $T_{\text{Joule}}$ for electron doping (Fig. 4b) is higher than for hole doping (Fig. 4c). As discussed below, this difference arises due to asymmetry in the conductivity (Fig. 4l) and in the inelastic scattering time $\tau_i$ of electrons and holes (Figs. 4n).

In contrast to the conventional Peltier effect in bulk materials, however, rather than showing a linear current dependence, $T_{\text{Pelt}}$ is described by a superposition of linear and cubic terms,

$$T_{\text{Pelt}} = a(I + bI^3), \tag{8}$$

as shown by the solid lines in Figs. 4e,f, where $a$ and $b$ are fitting parameters. The origin of the cubic term in $T_{\text{Pelt}}(I)$ can be understood as follows. As shown by Eq. 1, the Peltier coefficient depends quadratically on the electron temperature, $\Pi_{gr} = \tilde{\Pi}_{gr}T_e^2$. In a driven system, however, $T_e$ is current dependent. In fact, in a uniform case, this current dependence can be found by balancing Joule heating with the energy dumped to the phonons (Methods) yielding

$$T_e^2 = T_0^2 + \frac{\tau_i}{k_B^2 \nu(\varepsilon_F)\sigma}J^2 = T_0^2 + b'I^2, \tag{9}$$

where $T_0$ is the equilibrium temperature, $\tau_i$ is the inelastic scattering time, and $b'$ is a constant. Evaluating the terms in the transport equation which are responsible for the Peltier effect (Methods), we find

$$T_{\text{Pelt}}(I) \propto \dot{Q}_{\text{Pelt}}(I) = \tilde{\Pi}_j T_e^2 I = \tilde{\Pi}_j(T_0^2 I + b'I^3), \tag{10}$$

which motivates the observed current dependence of $T_{\text{Pelt}}(I) = a(I + bI^3)$. Note that the first and second terms reflect the linear and the nonlinear Peltier effects [49,50]. Importantly, Eqs. 9 and 10 are applicable only to a homogeneous system, and are presented here to provide insight into the origin of the nonlinear Peltier effects. Below, we derive the specific behavior at a p-n junction, in which Eqs. 9 and 10 will not be used.

The measured $|T_{\text{Pelt}}|$ decreases with doping $|n_c|$ (Figs. 4e,f), which is a direct consequence of the doping dependence of $|\tilde{\Pi}_j|$. This is also the reason for significantly larger $|T_{\text{Pelt}}|$ for p-n-p doping (Fig. 4e) as compared to p-p'-p (Fig. 4f) at comparable $|n_c|$. However, as discussed below, there is an additional contribution to this doping dependence originating from higher electron temperatures at low carrier densities due to higher Joule heating.

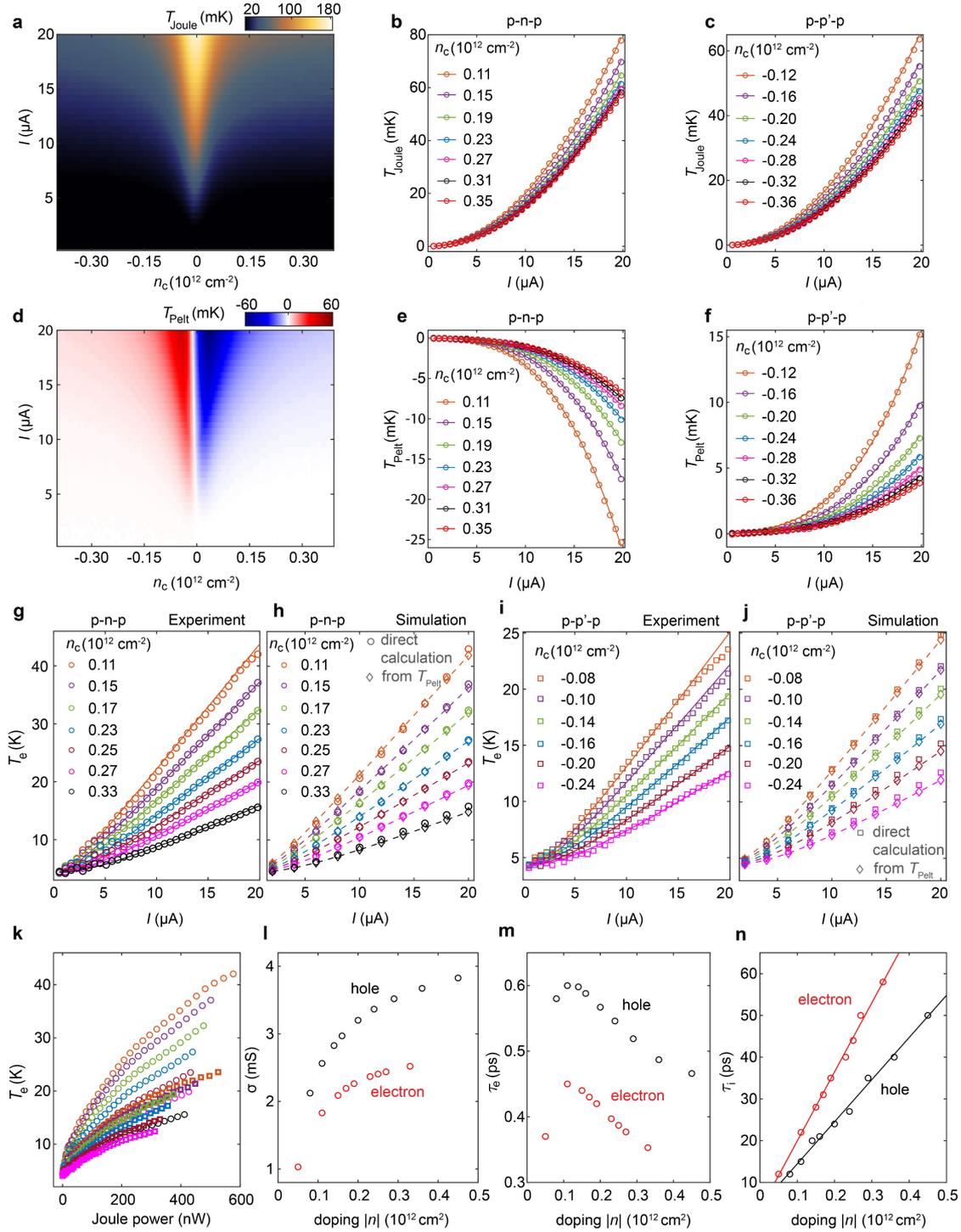

**Fig. 4. Nonlinear Peltier effect at graphene junction and derivation of the electron temperature. a,** $T_{Joule}$ at the right graphene junction measured at $T_0 = 4.3$ K as a function of bias current $I$ and carrier density in the central region $n_c$, while keeping the outer graphene regions hole doped at fixed $n_l = n_r = -0.43 \times 10^{12}$ cm$^{-2}$. **b,c,** $T_{Joule}$ vs. $I$ for various indicated electron (**b**) and hole (**c**) densities $n_c$ extracted from (a) (open circles). The solid curves are parabolic fit to the data $T_{Joule}(I) = cI^2$. **d,** $T_{Pelt}$ vs. $I$ and $n_c$ measured simultaneously with (a). **e,f,** $T_{Pelt}$ vs. $I$ for various indicated electron (**e**) and hole (**f**) doping $n_c$ extracted from (d). The solid curves are polynomial fit to the data, $T_{Pelt}(I) = a(I + bI^3)$. **g,i,** The rms electron temperature $T_e$ at the graphene junction vs. the rms current $I$ for various indicated electron (**g**) and hole (**i**) densities $n_c$. The open circles and squares in (g) and (i) show the $T_e(I)$ derived directly from $T_{Pelt}(I)$ data in (d) using Eq. 15, while the solid curves represent $T_e(I)$ derived from the polynomial fit to $T_{Pelt}(I)$. **h,j,** The simulated rms $T_e$ at the graphene junction vs. rms

current $I$ for various indicated electron (**h**) and hole (**j**) densities $n_c$. The open circles and squares in (h) and (j) show the rms $T_e$ derived directly from the electronic simulation, while the diamonds show $T_e$ calculated indirectly from COMSOL simulated surface lattice temperature $T_{Pelt}$, demonstrating the viability of the Peltier thermometry technique. **k,** The rms electron temperature $T_e$ at the graphene junction vs the total Joule power. The open circles and squares are the same symbols as in (g) and (i), respectively. **l,** Conductance of the graphene strip $\sigma$ vs. carrier density for electron (red) and hole doping (black). **m,** Elastic scattering time $\tau_e$ vs. carrier density derived from conductance in (k). **n,** Inelastic scattering time $\tau_i$ vs. carrier density derived from (h,j). The red and black lines are linear fits.

**Derivation of electron temperature at p-n junction**

Next, we show that by measuring the lattice temperature $T_{\text{Pelt}}(I)$, we can experimentally derive the electron temperature $T_e(I)$ at the p-n junction quantitatively. Let's first consider Joule heating. The generated heat $\dot{Q}(I)$ is dumped into the electron bath and then transferred to the lattice through some electron-phonon coupling mechanism. Since all the generated heat is eventually absorbed by the lattice, by measuring the Joule lattice temperature, one cannot determine the electron temperature: for weak electron-phonon coupling, the electron temperature $T_e(I)$ will be high, whereas for strong coupling, $T_e(I)$ will be low, but the corresponding lattice $T_{\text{Joule}}(I)$ will be hardly affected.

Now consider the Peltier effect. The measured lattice $T_{\text{Pelt}}(I)$ at a junction is given by Eq. 6,

$$T_{\text{Pelt}} = C\dot{Q}_{\text{Pelt}}^j = C\tilde{\Pi}_j T_e^2 I = C\tilde{\Pi}_j T_e^2(I) I, \tag{11}$$

The fundamental difference between $T_{\text{Joule}}(I)$ and $T_{\text{Pelt}}(I)$, is that the latter contains $T_e$ explicitly because the Peltier effect itself depends on the electron temperature. Let's define

$$T_e^2(I) = T_0^2 + f(I), \tag{12}$$

where the function $f(I)$ is unknown, except that $f(I=0) = 0$. Thus

$$T_{\text{Pelt}}(I) = C\tilde{\Pi}_j T_e^2(I) I = C\tilde{\Pi}_j T_0^2 I + C\tilde{\Pi}_j f(I) I, \tag{13}$$

$$T_e^2(I) = \frac{T_{\text{Pelt}}(I)}{C\tilde{\Pi}_j I}, \qquad C\tilde{\Pi}_j T_0^2 = \left.\frac{\partial T_{\text{Pelt}}(I)}{\partial I}\right|_{I=0}, \tag{14}$$

leading to our central result

$$T_e(I) = T_0 \sqrt{\frac{T_{\text{Pelt}}(I)/I}{\partial T_{\text{Pelt}}(I)/\partial I|_{I=0}}}. \tag{15}$$

Given the form of the measured $T_{Pelt}$, Eq. (8), we further obtain

$$T_e(I) = T_0\sqrt{1 + bI^2}, \tag{16}$$

where $b$ is an experimentally determined parameter. Thus, remarkably, measuring the nonlinear lattice Peltier temperature $T_{\text{Pelt}}(I)$ at a junction allows for direct determination of the electron temperature at the junction without knowledge of any material-dependent parameters. Equation 15 is a particular case of Eq. 7, in which the Joule heating and the Peltier effect are caused by the same current, resulting in the nonlinear $I$ dependence. In general, however, the heating can be induced by other means, in which case $T_{\text{Pelt}}(I)$ remains linear and $T_e$ is given by Eq. 7.

The open circles (squares) in Fig. 4g (4i) show the electron temperature $T_e(I)$ obtained directly from the measured individual $T_{\text{Pelt}}(I)$ points in Figs. 4e,f (open circles), using Eq. 15. The solid lines show a distinct derivation of $T_e(I)$ using the polynomial fit to $T_{\text{Pelt}}(I)$ data (solid lines in Figs. 4e,f) and Eq. 16. We find a very good agreement between the direct and polynomial derivations of $T_e$ for the various hole and electron doping levels $n_c$. Note that the derived $T_e$ reflects the rms temperature of the carriers at the junction without any knowledge of the microscopic electron heating and cooling mechanisms.

The obtained $T_e$ in Figs. 4g,i has several interesting observations. First is that at elevated currents, the electron temperature grows linearly with $I$ in contrast to the directly measured $T_{\text{Pelt}}$, which is cubic in $I$ (Figs 4e,f), and to $T_{\text{Joule}}$, which is quadratic in current (Figs 4b,c). This is the consequence of the fact that $T_e \gg T_0$ while $T_{\text{Joule}} \ll T_0$. Depending on the doping level, the current-induced electron heating ratio is in the range of 0.5 to 3 K/µA for our hBN encapsulated graphene strip of width $W = 3.1$ µm at 4.3 K. An additional observation is that $T_e$ is significantly higher for the p-n-p configuration (Fig. 4g) as compared to p-p'-p (Fig. 4i) for comparable doping levels. This asymmetry is attributed to two factors. First, the resistance for electron doping is higher than for hole doping (Fig. 4l and dashed line in Fig. 3d), leading to larger Joule heating. Second, the inelastic scattering time ($\tau_i$) for electrons is longer than for holes (Fig. 4n), resulting in a lower electron cooling rate. The higher $T_e$ for electrons further leads to a larger $|T_{\text{Pelt}}|$ as evident from Eq. 10 and observed in Figs. 4e,f. We note that all the presented data are for carrier densities away from CNP where the I-V characteristics are linear. Very close to CNP the presented analysis becomes inapplicable due to nonlinear I-V characteristics arising from self-gating effects [1,5].

The $T_e$ results in Figs. 4g,I are presented in Fig. 4k vs. the total Joule power dissipated in the sample. Power as low as 10 nW raises the electron temperature by several K as reported previously [4,32]. Moreover, the same heating power results in very different $T_e$ at different dopings due to strong dependence of $\tau_i$ on carrier density (Fig. 4n and Methods).

**Numerical simulations**

To gain additional insight into the Peltier electron thermometry, we performed two-step numerical simulations for our device geometry. In the first step, the non-equilibrium electron temperature profile $T_e(x)$ in the graphene strip is calculated using semiclassical transport equations in the presence of an *ac* current with an inelastic scattering time $\tau_i$ governing the energy relaxation rate to the lattice (Eq. M6 in Methods). The resulting profile of the energy relaxation rate along with the $\dot{Q}_{\text{Pelt}}$ contribution are then used as a source term for solving 3D heat diffusion equations in the encapsulating hBN structure and in Au contacts using COMSOL. 2D maps of $T_{\text{Pelt}}$ and $T_{\text{Joule}}$ are then derived as the first and second harmonic components of the *ac* lattice temperature at the surface of the device, respectively. The obtained $T_{\text{Pelt}}$ and $T_{\text{Joule}}$ maps for the uniformly hole-doped graphene strip are presented in Figs. 1h,k, which match very well both qualitatively and quantitatively with the measured maps in Figs. 1f,j. In particular, the Peltier effect and the elevated $T_{\text{Joule}}$ at the Au-graphene junctions are well reproduced along with the enhanced heat conductivity along the extended Au contacts. A similarly good fit is obtained for uniform electron doping as shown in Figs. 1g,i for $T_{\text{Pelt}}$ and in Extended Data Figs. 2a,c for $T_{\text{Joule}}$.

Figures 2e,f and Extended Data Figs. 3a,b show that the simulated $T_{\text{Pelt}}$ for the p-n-p and n-p-n configurations in graphene strip are in close correspondence with the experiment. The simulated maps readily reproduce the much stronger Peltier heating and cooling at the respective p-n junctions as compared to Au/graphene junctions, as expected from the larger $|\tilde{\Pi}_j|$ and the higher $T_e$ in the bulk of graphene than at the Au contacts. The numerically derived $T_e(x)$ in graphene in the various configurations is presented in Extended Data Figs. 6 to 8. Notably, most of the electron heating arises from Joule dissipation, while the Peltier effect locally modifies the electron temperature in the vicinity of the p-n junctions.

The two-step simulations allow direct testing of the validity of the Peltier electron thermometry principle. The open circles in Figs. 4h,j show the electron temperature $T_e(I)$ at the graphene junction for various doping levels calculated directly from transport equations, in which the electrons couple to the environment only through the inelastic scattering time $\tau_i$. In contrast, the open diamonds show $T_e(I)$ derived from the surface lattice temperature $T_{\text{Pelt}}$ calculated in the second step using Eq. 15. In this step, 3D heat diffusion equations are solved outside the graphene having only the energy relaxation rate as an input with no direct information on $T_e$. The agreement between the two results is striking, corroborating the validity of the developed technique. We emphasize that such derivation of $T_e$ from the even-harmonic $T_{\text{Joule}}$ is not possible, unlike the odd-harmonic $T_{\text{Pelt}}$, the magnitude of which is explicitly determined by $T_e$.

From fitting the simulated $T_e$ in Figs. 4h,j to the experimental data in Figs. 4g,l, we extract $\tau_i$ (Eq. 9 and methods) for various carrier densities (Fig. 4n), which is found to increase monotonically with both electron and hole doping, revealing a decrease in the electron-phonon cooling power with doping (Eq. M6, Methods). We attribute the inelastic scattering mechanism to impurity-assisted electron-phonon scattering, predominantly governed by resonant states due to atomic defects at graphene edges [48,51]. The density of these states has been shown to decrease away from charge neutrality [48], consistent with the observed increase of $\tau_i$ with $|n|$. Moreover, the width of the resonant levels grows with $|n|$ [52], causing an additional rise in $\tau_i$ with doping. The observed asymmetry in $\tau_i$ between the electron and hole doping is likely due to the type of defects creating the resonant states below and above the Dirac point [52]. Note that at high dopings, $\tau_i$ is some two orders of magnitude larger than the elastic scattering time $\tau_e$ (Fig. 4l) derived from the measured conductance $\tau_e = \frac{h}{e^2}\frac{\sigma}{2k_F v_F}$, where $k_F = \sqrt{\pi n}$ and $v_F$ is the Fermi velocity.

**Discussion**

We have introduced a novel approach to electronic thermometry that utilizes nanoscale imaging of the Peltier effect at cryogenic temperatures. This Peltier thermometry method, which exploits local cooling and heating at electrostatically defined p–n junctions in graphene, provides a minimally invasive and highly sensitive probe of the electronic temperature. Its utility extends broadly across low-dimensional systems, especially those where electronic transport is tightly coupled with thermal processes.

One particularly promising application is graphene moiré systems, where current bias can induce puzzling collective behavior originating from multiband electronic structures and strong interactions [7,53]. These phenomena, which reflect rich nonequilibrium dynamics, are ideally suited for investigation with nanoscale thermometry. In standard Hall bar geometries, gate-defined p-n junctions can be embedded into the side contact regions to locally extract electron temperature at multiple points without perturbing the main transport channel, offering powerful insight into how energy flow and dissipation drive the formation of emergent ordered phases. Similar approaches may also be used to probe spatially resolved heat transport in other strongly interacting electron systems or hydrodynamic electron fluids.

Beyond its broad applicability, Peltier thermometry enables access to microscopic cooling mechanisms that are otherwise difficult to probe directly. Our separate access to the Peltier and Joule signals allows extraction of the temperature dependence of the cooling power from their current dependence. Figures 4e,f show that the measured Peltier signal has pure $I^3$ dependence at high currents, which confirms the $T_e^2$ dependence of the cooling power as shown in Extended Data Fig. 5 and discussed in Methods. Strikingly, our finding that the cooling power scales as $T_e^2$ and decreases with increasing carrier density departs significantly from existing theoretical models (see [2] and references therein), pointing to significant gaps in our current understanding of electronic cooling pathways in graphene. Although $T_e^2$ scaling has been predicted for momentum-conserving two-phonon processes [2], this model does not capture the observed doping dependence. This discrepancy indicates the likely involvement of additional mechanisms such as interband polarization and screening or, possibly, impurity-assisted scattering by localized states at graphene flake edges [48]. In this scenario, the extracted dependence of the inelastic scattering time $\tau_i \propto |n|$ provides a window into the spectral properties of those resonant states and their evolution with doping. These insights highlight the need for further theoretical and experimental exploration of edge-related cooling pathways and electron–phonon interactions in two-dimensional materials.

Importantly, our measurements show that the lattice temperature increase from Joule heating $T_{\text{Joule}}$ (Figs. 4b,c) consistently exceeds that from Peltier cooling $T_{\text{Pelt}}$ (Figs. 4e,f), indicating that only relative (not absolute) cryogenic cooling is achieved under our conditions. This contrasts with recent reports of absolute cooling at 4 K in WTe₂ via the Ettingshausen effect [31]. However, we uncover a distinct nonlinear thermoelectric regime in graphene: at elevated electron temperatures, the Peltier effect becomes strongly nonlinear [49], scaling as $I^3$, while Joule heating grows only as $I^2$. This difference in scaling suggests that, under suitable conditions, nonlinear Peltier cooling could overcome Joule heating—potentially enabling absolute cooling at cryogenic temperatures in the regime where conventional thermoelectric cooling effects are suppressed.

**Acknowledgments**

This work was co-funded by the United States - Israel Binational Science Foundation (BSF) grant No 2022013, by the Israel Science Foundation ISF grant No 687/22, and by the European Union (ERC, MoireMultiProbe - 101089714). Views and opinions expressed are however those of the author(s) only and do not necessarily reflect those of the European Union or the European Research Council. Neither the European Union nor the granting authority can be held responsible for them. S.K.S. acknowledges the Israel Academy of Sciences and Humanities (IASH) and Council for Higher Education (CHE) Excellence Fellowship. S.K.S. further acknowledges the financial support from the Weizmann postdoctoral Excellence Fellowship. E.Z. acknowledges the support of the Goldfield Family Charitable Trust, the Andre Deloro Prize for Scientific Research, and Leona M. and Harry B. Helmsley Charitable Trust grant #2112-04911. The work of D.A.P. was supported by the National Science Foundation Grant No. DMR-2138008. K.W. and T.T. acknowledge support from the JSPS KAKENHI (Grant Numbers 21H05233 and 23H02052), the CREST (JPMJCR24A5), JST and World Premier International Research Center Initiative (WPI), MEXT, Japan.

**Author contributions**

S.K.S., T.V., and E.Z. designed the experiment. S.K.S. fabricated and characterized the samples. S.K.S and T.V. performed the local thermal imaging measurements. S.K.S. and T.V. performed the analysis. T.V., S.K.S., and Y.M. fabricated the JOT and tuning fork. M.E.H. developed the JOT readout. G.Q., L.S.L., and D.A.P., performed the theoretical modelling. G.Q. performed the numerical modeling of transport equations. S.K.S. performed the finite element COMSOL numerical simulations. S.K.S., T.V., G.Q., L.S.L., D.A.P, and E.Z. wrote the manuscript. K.W. and T.T. provided the hBN crystals.

**Competing interests**

The authors declare no competing interests.

**Data availability**

The data that support the findings of this study are available from the corresponding authors on reasonable request.

**Code availability**

The transport equations and finite-element COMSOL numerical simulation codes are available from the corresponding author on reasonable request.

# Methods

## Device fabrication

The hBN encapsulated monolayer graphene stack was fabricated using the standard dry-transfer pick-up technique with graphene and hBN flakes exfoliated onto a $SiO_2$(285 nm)/Si surface. The exfoliated flakes were picked up using a polycarbonate (PC) coated polydimethylsiloxane (PDMS) dome stamp attached to a micromanipulator. The resulting stack was transferred onto a prefabricated local Pt gate structure at 180 °C. The finalized stack was vacuum annealed at 350 °C to minimize nano-bubbles and blisters in the stack. 1D edge contacts were defined by reactive ion etching in a mixture of $CHF_3$ and $O_2$ gases with flow rates of 40 sccm and 4 sccm, respectively at 25 °C with RF power of 60 W. The etching time was optimized such that the hBN below the graphene flake did not etch completely to isolate the contacts from the bottom metal gate structures. Finally, Cr/Au (3/80 nm) was deposited to achieve the 1D metal/graphene edge contacts.

The three metal back gates were fabricated as follows. First a Ti (2nm)/Pt (10 nm) global gate for doping the central region of the graphene strip was deposited onto a $SiO_2$/Si surface. After the deposition and liftoff procedures, the metal gate structure was vacuum annealed at 450 °C. Then, a hBN flake was picked up using a PC coated PDMS stamp and transferred onto this pre-patterned Ti/Pt gate. The resulting structure was cleaned in chloroform, followed by vacuum annealing at 450 °C. Next, two local Ti (2nm)/Pt (8 nm) strips for doping the right and left graphene regions were deposited on top of the fabricated hBN/Pt/Ti/$SiO_2$/Si structure and were again vacuum annealed at 450 °C.

## JOT fabrication and thermal imaging

SQUID on tip (SOT) has been shown to provide highly sensitive thermal and magnetic nanoscale imaging [48,51,54–60]. Since SQUID magnetic imaging is based on interference, it requires two superconducting junctions. Thermal imaging in contrast, is mainly based on the suppression of the critical current with temperature [29], and can thus be achieved with a single junction. Since in this work, we are interested only in thermal imaging, we have developed a single junction on a tip (JOT) device, which simplifies the device fabrication and allows attaining smaller apex diameters. A grooved quartz rod was pooled to a sharp needle, followed by three step deposition of MoRe using a collimated sputtering technique as developed for SOT fabrication [61], forming a single superconducting constriction at the apex (Extended Data Fig. 1).

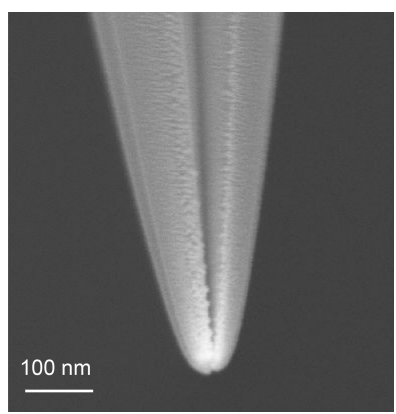

**Extended Data Fig. 1. Scanning electron microscope image of the junction-on-tip (JOT).** The JOT is fabricated by three-step deposition of MoRe on a pulled grooved quartz rod with apex diameter of about 50 nm.

In this work, a MoRe JOT with a diameter of about 50 nm and thermal sensitivity of 2.6 μK/$Hz^{1/2}$ was used as a scanning nanoscale thermal sensor. The sample chamber was filled with low-pressure He exchange gas to provide the required thermal coupling between the JOT and the sample surface. Lownoise readout of the JOT signal was achieved by using a SQUID series array amplifier [62]. To control the scan height, the JOT was mechanically attached to a quartz tuning fork vibrating at a resonant frequency of about 32.8 kHz as described in [29]. Thermal imaging was performed at a constant scan height of 200 nm above the sample surface at a base temperature of 4.3 K and He exchange gas pressure of about 25 mbar. At this pressure, the JOT is in thermal equilibrium with the local surface temperature of the sample [29]. The presented 2D scans were acquired with image resolution of 180×53 pixels with pixel size of 130 nm and acquisition time of 40 ms per pixel.

Extended Data Fig. 2 shows the measured and simulated Joule lattice temperature $T_{\text{Joule}}$ for doping configurations opposite to those presented in Figs. 1j,k and 2g,h.

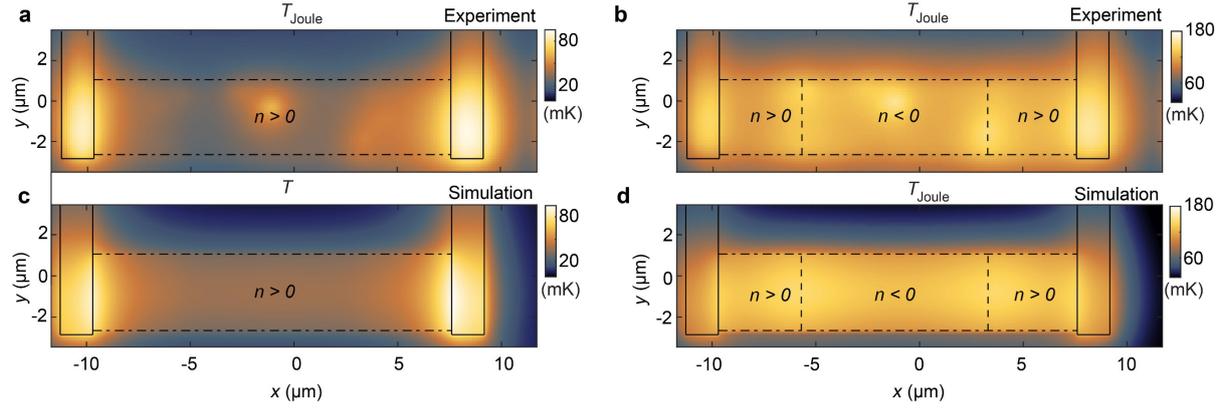

**Extended Data Fig. 2. Maps of $T_{\text{Joule}}(x, y)$ for uniform and n-p-n doping configurations. a,** $T_{\text{Joule}}$ map acquired with $I = 20$ µA rms at $T_0 = 4.3$ K for the uniform electron doping of graphene with $n = 0.27 \times 10^{12}$ cm$^{-2}$. The small hot spots in the center and in the right regions of the device are caused by enhanced electron-phonon scattering at bubble defects in graphene structure visible in the optical image in Fig. 1c. **b,** Sama as (a) for n-p-n doping at $(0.08, -0.29, 0.08) \times 10^{12}$ cm$^{-2}$. **c,d,** Numerically simulated $T_{\text{Joule}}$ maps of the device for the same current and doping values as in a,b.

An additional example of measured and simulated $T_{Pelt}$ and $T_{Joule}$ maps for the case of p-n-p doping with high carrier concentrations in the left and right regions and low doping in the central region is shown in Extended Data Fig. 3.

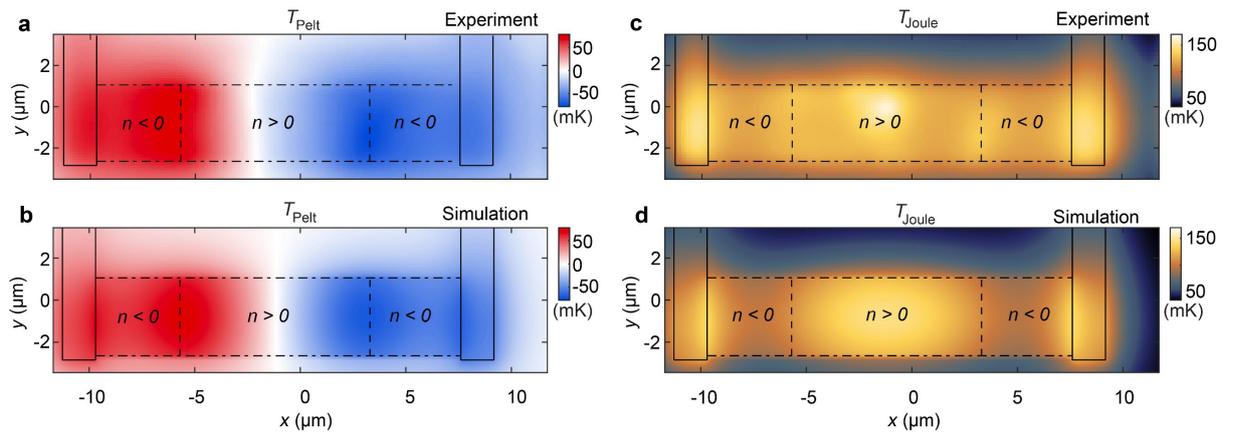

**Extended Data Fig. 3. $T_{Pelt}$ and $T_{Joule}$ maps for additional doping configurations. a,** $T_{Pelt}$ map acquired for $I = 20$ µA rms at $T_0 = 4.3$ K for the p-n-p doping of $(-0.45, 0.05, -0.45) \times 10^{12}$ cm$^{-2}$. **b,** Numerically simulated $T_{Pelt}$ across the device for the same current and doping values as in (a). **c,** Joule lattice temperature $T_{Joule}$ acquired simultaneously with (a). The small hot spot in the center of the devices is caused by enhanced electron-phonon scattering at a bubble defect in graphene structure. **d,** Numerically simulated $T_{Joule}$ for the same conditions as in (c).

To provide a qualitative comparison to the measured $T_\text{Pelt}$ presented in Fig. 3c, Extended Data Fig. 4 shows $\dot{Q}_\text{Pelt}$ calculated from the measured two-probe resistance of the device using Eq. 1.

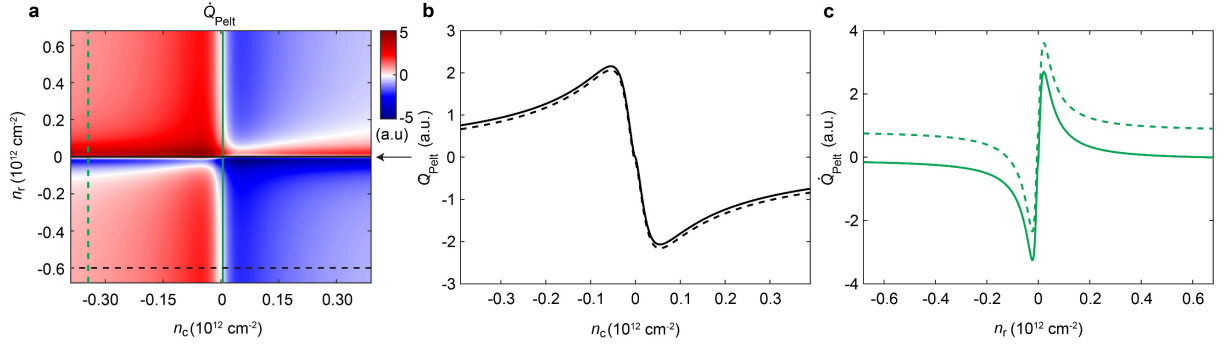

**Extended Data Fig. 4. Calculated $\dot{Q}_\text{Pelt}$ from two-probe resistance of the central and right channels. a,** $\dot{Q}_\text{Pelt}$ map at the right p-n junction calculated from the two-probe resistance response of central $n_c$ and outer right-side $n_r$ graphene regions. **b,** Linecuts of the calculated $\dot{Q}_\text{Pelt}$ along the black solid and dashed lines in (a) at $n_r = 0$ and $-0.6 \times 10^{12}$ cm$^{-2}$. **c,** Linecuts of $\dot{Q}_\text{Pelt}$ along the green solid and dashed lines in (a) at $n_c = 0$ and $-0.35 \times 10^{12}$ cm$^{-2}$.

**Heat transport equations**

Below we formulate and solve the thermal transport problem in a graphene strip in the presence of both transport electric current and electronic temperature gradients. The existence of the electronic temperature is the underlying assumption of the theory. However, even in strongly nonequilibrium situations, when the temperature of the electronic system is undefined, we can still think of $T_e$ as an effective quantity that determines the local energy density, so the equations below should be qualitatively correct.

We further assume that the phonon bath, with which electrons exchange energy in the process of their cooling, always remains close to the base temperature as directly corroborated by our experimental observations. For the theory developed, this implies that the transport equation for the electronic subsystem only contains coupling to a cold reservoir at constant temperature. The equation for the phonon subsystem can then be solved independently, treating the electrons as an external heat source, the intensity of which depends on the local electronic and base temperatures.

Consistent with the assumption of the existence of local electronic temperature, we assume that deviations from local equilibrium are small. However, the local equilibrium can be characterized by a temperature very different from the base one. The expressions for the electric, $J$, and electron carried heat, $J_Q$, current densities to linear order in gradients are

$$J = -\sigma\left(\frac{\nabla \mu_{ec}}{e} + S \nabla T_e\right), \tag{M1}$$

$$J_Q = -\kappa \nabla T_e + \Pi J. \tag{M2}$$

These expressions define the transport coefficients in the electron system, which themselves can be nonlinear functions of temperature and density. Here $\mu_{ec}$ is the electrochemical potential, $S$ and $\Pi$ are the Seebeck and Peltier coefficients, $\Pi = S T_e$, which should still hold in local equilibrium. We assume that $\mu_{ec}$ and $T_e$ entering the transport currents are continuous throughout the sample, which implies that various junctions have small electric and heat resistances as compared to that of the sample.

Equation M2 for the heat current density explicitly contains the contribution from electric current via the Peltier effect. When there are discontinuities in the Peltier coefficient $\Pi$ across junctions, the continuity of the electric current implies that the Peltier component of the heat current becomes discontinuous. This discontinuity gives rise to localized Peltier heating or cooling at the junctions, as described by $\dot{Q}_{\text{Pelt}}^j$ in Eqs. 2 and 3 of the main text.

Assuming steady-state regime with a uniform electric current, the heat transport equation reads as

$$-\nabla(\kappa \nabla T_e) + T_e J \cdot \nabla S = \frac{J^2}{\sigma} + P_{e-ph}(T_e, T_0). \tag{M3}$$

The first term on the left-hand side is the divergence of the conventional heat current in the absence of thermoelectric effects. The second term contains the Peltier effect at junctions, where $\tilde{\Pi} = S/T_e = \Pi/T_e^2$ changes discontinuously, as well as the Thomson effect due to the smooth coordinate dependence of the Seebeck coefficient away from junctions due to nonuniform temperature profile, $S(x) = S(T_e(x))$. Finally, the right-hand side of the equation contains the Joule heating (the first term), as well as the cooling power $P_{e-ph}$ of electrons due to emission of acoustic phonons, the form of which is discussed below.

The transport coefficients $\kappa, \sigma, \Pi$ are functions of local temperature and density. In general, they are phenomenological and can only be determined fully by experiment. However, a simple model can be obtained by extracting them from the linear response theory for free electrons, given values of 'equilibrium' $T_e, \mu_{ec}$. This yields

$$\kappa = \tilde{\kappa} T_e, \quad \Pi = \tilde{\Pi} T_e^2, \tag{M4}$$

where $\tilde{\kappa}$ and $\tilde{\Pi}$ are doping-dependent material constants. Note that the resulting nonlinear transport equations have validity beyond the linear response used to obtain the coefficients. The temperature-independent reduced heat conductivity $\tilde{\kappa}$ is inferred from the Wiedemann-Franz law:

$$\tilde{\kappa} = \frac{\pi^2 k_B^2}{3e^2} \sigma, \tag{M5}$$

while the reduced Peltier coefficient is given by Eq. (1), which we repeat here for convenience:

$$\tilde{\Pi} = \frac{\pi^2 k_B^2}{3e} \frac{\partial \ln \sigma}{\partial n} \nu(\varepsilon_F).$$

The electrical conductivity $\sigma$, which enters the expressions for $\tilde{\kappa}$ and $\tilde{\Pi}$ is measured experimentally.

It is customary to choose the cooling power as $P_{e-ph}(T_e, T_0) = -\alpha(T_e^\delta - T_0^\delta)$ with constants $\alpha$ and $\delta$, that depend on active cooling pathways. For instance, for momentum-conserving electron phonon scattering $\delta = 4$ below the Bloch-Gruneisen temperature, while $\delta = 1$ above it [63–65]. Existing models for impurity-assisted supercollisions predict $\delta = 3$ or $5$ [52,66,67]. In all these cases the prefactor $\alpha$ is complicated and is not shown here. It is apparent from the theoretical results of Refs. [52,67] and experimental results of Ref. [48] that the specific values of $\alpha, \delta$ may depend rather sensitively on the type of disorder in the system. In this work we choose a cooling power that best fits the experimental data at low doping, where we expect electron-phonon scattering to play the most significant role in electron cooling. The proposed form of the electron-phonon cooling power is

$$P_{e-ph}(T_e, T_0) = -\frac{k_B^2 \nu(\varepsilon_F)}{\tau_i}(T_e^2 - T_0^2). \tag{M6}$$

As before, $\nu(\varepsilon_F)$ denotes the density of states at the Fermi level, which for graphene is given by $\nu(\varepsilon_F) = \frac{2}{\hbar v_F}\sqrt{\frac{|n|}{\pi}}$. For continuity of presentation we defer the justification of Eq. M6 till the end of this section. Here we only note that Eq. M6 can be phenomenologically derived from the constant relaxation time approximation for the electron-phonon collision integral, but, again, its general form is motivated by our experimental results. In principle, the expression for the cooling power can be phenomenologically generalized to have a $T_e$-dependent $\tau_i$. We did not need to do so to achieve good fits to experimental data.

The aforementioned considerations bring the transport equation M3 to the following form:

$$-\frac{1}{2}\nabla(\tilde{\kappa}\nabla T_e^2) + \frac{1}{2}J\tilde{\Pi}\cdot\nabla T_e^2 + JT_e^2\nabla\tilde{\Pi} = \frac{J^2}{\sigma} - \frac{k_B^2\nu(\varepsilon_F)}{\tau_i}(T_e^2 - T_0^2). \tag{M7}$$

It is important to note that the expressions for the transport currents that led to this steady-state transport equation were derived assuming small deviations from local equilibrium, and therefore the fluxes are linear in the electric field and temperature gradient. However, the local equilibrium value of the electronic temperature itself depends on the value of the electric current because of the Joule heating and thermoelectric phenomena. Therefore, the resulting transport equations describe nonlinear response to the transport electric current. Another feature of the derived transport equation is its linearity in $T_e^2$. This linearity can easily break down if one, say, uses a more complicated form of the electron-phonon cooling rate, or more involved expressions for the transport coefficients. However, it allows for a greatly simplified analysis which is insensitive to details in the experimental setup, and even the current form of the transport equation M7 leads to a good agreement with the experimental data.

Equation M7 can be written more compactly as

$$\nabla(-\tilde{\kappa}T_e\nabla T_e + \beta T_e^2) = \beta T_e\nabla T_e + \Sigma - \Omega(T_e^2 - T_0^2), \tag{M8}$$

where $\Sigma = \frac{J^2}{\sigma}$, $\Omega = \frac{k_B^2\nu(\varepsilon_F)}{\tau_i}$, and $\beta = J\tilde{\Pi}$. Assuming that in each doping region the coefficients are constant and $T_e$ varies only along $x$, a linear differential equation in the square of the temperature $t(x) = T_e^2(x)$ is obtained,

$$-\frac{\tilde{\kappa}}{2}t'' + \frac{1}{2}\beta t' = \Sigma - \Omega(t - t_0), \tag{M9}$$

which has a general solution

$$t(x) = C_1 e^{\frac{x(\beta/2 - \sqrt{2\Omega\tilde{\kappa} + \beta^2/4})}{\tilde{\kappa}}} + C_2 e^{\frac{x(\beta/2 + \sqrt{2\Omega\tilde{\kappa} + \beta^2/4})}{\tilde{\kappa}}} + \frac{\Sigma}{\Omega} + t_0. \tag{M10}$$

Note that the uniform solution is given by

$$t = t_0 + \frac{\Sigma}{\Omega},$$

corresponding in the original quantities to

$$T_e^2 = T_0^2 + \frac{\tau_i}{k_B^2\nu(\varepsilon_F)\sigma}J^2, \tag{M11}$$

which is precisely what one gets by balancing the Joule heating with the heat dumped into the phonons. The boundaries between different regions can be matched by demanding both the temperature $t(x)$ and the in-plane heat flux, $\dot{Q}(x) = -\frac{\tilde{\kappa}}{2}t'(x) + \beta t(x)$, expressed by the left-hand side of Eq. M8, to be continuous across the junctions. The only remaining conditions for deriving the temperature profile are the boundary conditions at the right and left graphene-Au contacts. Since the electron-phonon coupling in the metal is strong and since the Au leads act as a good heat sink for the phonons, it is reasonable to assume that $T_e$ at the graphene-Au contact will approach the base temperature, $t_0 = T_0^2$. Thus, the $T_e$ solution along the graphene strip can be derived without reference to the rest of the system. We can then use the resulting $T_e(x)$ to source the COMSOL 3D simulation as heat driven into the substrate in the $z$ direction at each point along the graphene according to Eq. M6. In addition, to account for the heat dumped by the electrons into the leads, we calculate the lateral electron heat fluxes at the right and left boundaries and input them as line heat sources into the Au leads in COMSOL.

We end this Section with a discussion of our choice of $\delta = 2$ in the expression for the cooling power, $P_{e-ph}(T_e, T_0) = -\alpha(T_e^\delta - T_0^\delta)$. This form makes Eq. M8 readily solvable numerically because it is a linear differential equation in $T_e^2$ as described by Eq. M9.

We assume that at the lowest doping treated in this work, $|n| \sim 10^{11}$ cm$^{-2}$, electron-phonon scattering determines the steady-state value of the electronic temperature throughout most of the sample. Only within short regions near the contacts energy diffusion into the contacts is important to bring $T_e$ down to the base temperature as seen in Extended Data Figs. 6b,h. For a qualitative evaluation, we can then analyze a uniform sample with electron cooling occurring only through the electron-phonon channel. In this case, Eq. M3 reduces to

$$I^2 R = \alpha(T_e^\delta - T_0^\delta), \tag{M12}$$

where $R$ is the total resistance of the sample and $\alpha$ is a constant, with a solution

$$T_e = \left(\frac{I^2 R}{\alpha} + T_0^\delta\right)^{1/\delta}. \tag{M13}$$

Consequently, the Peltier temperature is given by

$$T_{\text{Pelt}}(I) = \beta T_e^2 I = \beta \left(\frac{I^2 R}{\alpha} + T_0^\delta\right)^{\frac{2}{\delta}} I, \tag{M14}$$

where $\beta$ is a constant. In the high-current limit, $T_{\text{Pelt}}$ reduces to

$$T_{\text{Pelt}}(I) = \frac{\beta R}{\alpha} I^{\left(\frac{4}{\delta}+1\right)}. \tag{M15}$$

Extended Data Fig. 5 shows the measured $T_{\text{Pelt}}(I)$ from Fig. 4e along with the fits to Eq. M14 for different values of $\delta$. While $\delta = 2$ shows an excellent fit, all the higher orders fail to reproduce the experimental current dependence. The reason is that at elevated currents all the results in Figs. 4e,f show $T_{\text{Pelt}}(I) \propto I^3$, which means $\delta = 2$ in Eq. M15. The higher $\delta$, in contrast, result in current dependences that are much weaker than $I^3$, with powers of 7/3, 2, 9/5 for $\delta = 3, 4, 5$, respectively.

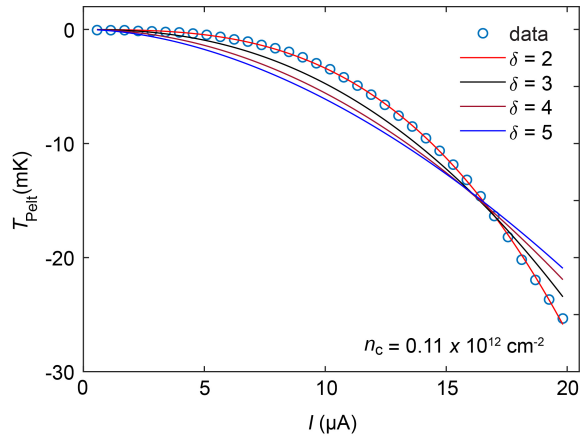

**Extended Data Fig. 5. Deriving the exponent of electron-phonon cooling power.** $T_{\text{Pelt}}$ (open circles) vs. $I$ for p-n-p doping of $(-0.43, 0.11, -0.43) \times 10^{12}$ cm$^{-2}$. The solid curves are the fit to the data with $T_{\text{Pelt}}(I) = \beta(\alpha I^2 + T_0^\delta)^{2/\delta} I$ for various indicated exponents $\delta$. The best fit to experimental data is obtained for $\delta = 2$.

Having established the temperature dependence of the electron-phonon cooling power, we can express the coefficient $\alpha$ phenomenologically as $\alpha = k_B^2 \nu(\varepsilon_F)/\tau_i$, as in Eq. M6. The parameter $\tau_i$ then plays the role of the inelastic scattering time in the transport equation M7.

To discuss the consistency of the laid out physical picture at low doping, we note that Eq. M7 suggests that electron energy diffusion into contacts can also be considered as an effective cooling power that is linear in $T_e^2$. It can be roughly thought of in terms of Eq. M6 with the inelastic relaxation time replaced with the diffusion time along a uniformly doped part of the sample, $\tau_{\text{diff}} \sim \frac{L^2}{v_F^2 \tau_e}$, where $L$ is the length of that part. This leads us immediately to the conclusion that the $I^3$ dependence of $T_{\text{Pelt}}$ at high bias current should persist for all doping levels, which is what we observe experimentally.

Finally, we stress that despite the $I^3$-dependence of $T_{\text{Pelt}}$ at any doping regardless of the dominant cooling power – electron-phonon or diffusion into contacts – our assumption that its magnitude is determined by electron-phonon scattering at low doping is justified *a posteriori* by numerical modeling, see below. By fitting the measured profiles of $T_{\text{Pelt}}$ and $T_{\text{Joule}}$, we extract the inelastic time dependence on density as presented in Fig. 4n. It is then seen that for $|n| \sim 10^{11}$ cm$^{-2}$ we get $\tau_i \sim 10$ ps. The diffusion time along, say, the central part of the sample can be obtained using the elastic transport time extracted from the conductivity measurements (Fig. 4m) and is almost density-independent: $\tau_{\text{diff}} \sim 150$ ps, such that $\tau_{\text{diff}} \gg \tau_i$. This shows that our assumption of electron-phonon cooling power being the dominant cooling pathway for a uniform sample with low doping is self-consistent.

**Numerical simulations**

Below we use $P_{e-ph}(T_e, T_0)$ given by Eq. M6 to perform numerical simulations in two steps. First, we solve Eq. M7 to get the electron temperature profile along the graphene strip and the corresponding heat flux dumped into the lattice described by Eq. M6. In the second step, the first- and second-harmonic components of the dumped heat flux are used to solve 3D heat transport equations in our specific device geometry by finite element 3D simulations using COMSOL Multiphysics 6.2. The 3D simulation geometry includes the graphene layer, top hBN (35 nm), bottom hBN (58 nm) below the graphene layer, local Pt gates (10 nm), bottom-most hBN (52 nm) between the Pt local gate and the global Pt gate (12 nm), and a He exchange gas layer surrounding the sample extending 1 μm above the sample surface. The SiO$_2$ and Si layers were not included in the simulations because we set the bottom surface of the global Pt gate to bath temperature, $T_0 = 4.3$ K. We then find the 2D simulated distributions of $T_{\text{Pelt}}$ and $T_{\text{Joule}}$ that match well with the experiment as shown in Figs. 1, 2, and 4 and Extended Data Figs. 2, and 3 using material parameters as described below. After setting the material properties, the only remaining doping-dependent free fitting parameter is $\tau_i$ in Eq. M6, which we determine by fitting the simulated $T_e$ at the junction in Figs. 4h,j to the experimental values in Figs. 4g,i. The resulting $\tau_i$ values are presented in Fig. 4n.

To gain further understanding, Extended Data Fig. 6a shows $T_e(x)$ calculated from Eq. M7 for a *dc* current of $I_{dc} = 20$ μA applied from left to right (solid blue line) in a p-n-p configuration with high doping in the outer regions and low doping in the center, with carrier densities $(-0.45, 0.05, -0.45) \times 10^{12}$ cm$^{-2}$ and $\tau_i$ values of $(50, 12, 50)$ ps, corresponding to parameters in Extended Data Fig. 3. For positive $I_{dc}$ the left p-n junction is in forward bias resulting in Peltier heating at the junction, whereas the right n-p junction is reverse biased causing Peltier cooling. The profile has opposite shape for $I_{dc} = -20$ μA (dashed). Extended Data Fig. 6b presents the rms value of $T_e(x)$ in presence of an *ac* current showing a maximum in the center, a small incline at the two junctions, and $T_e = T_0$ at the right and left graphene edges that are in contact with Au electrodes. The first harmonic of $T_e$, which is much smaller than the rms value, reflects the thermoelectric response in the electron bath (Extended Data Fig. 6c), showing peaks in Peltier heating and cooling at the left and right junctions respectively.

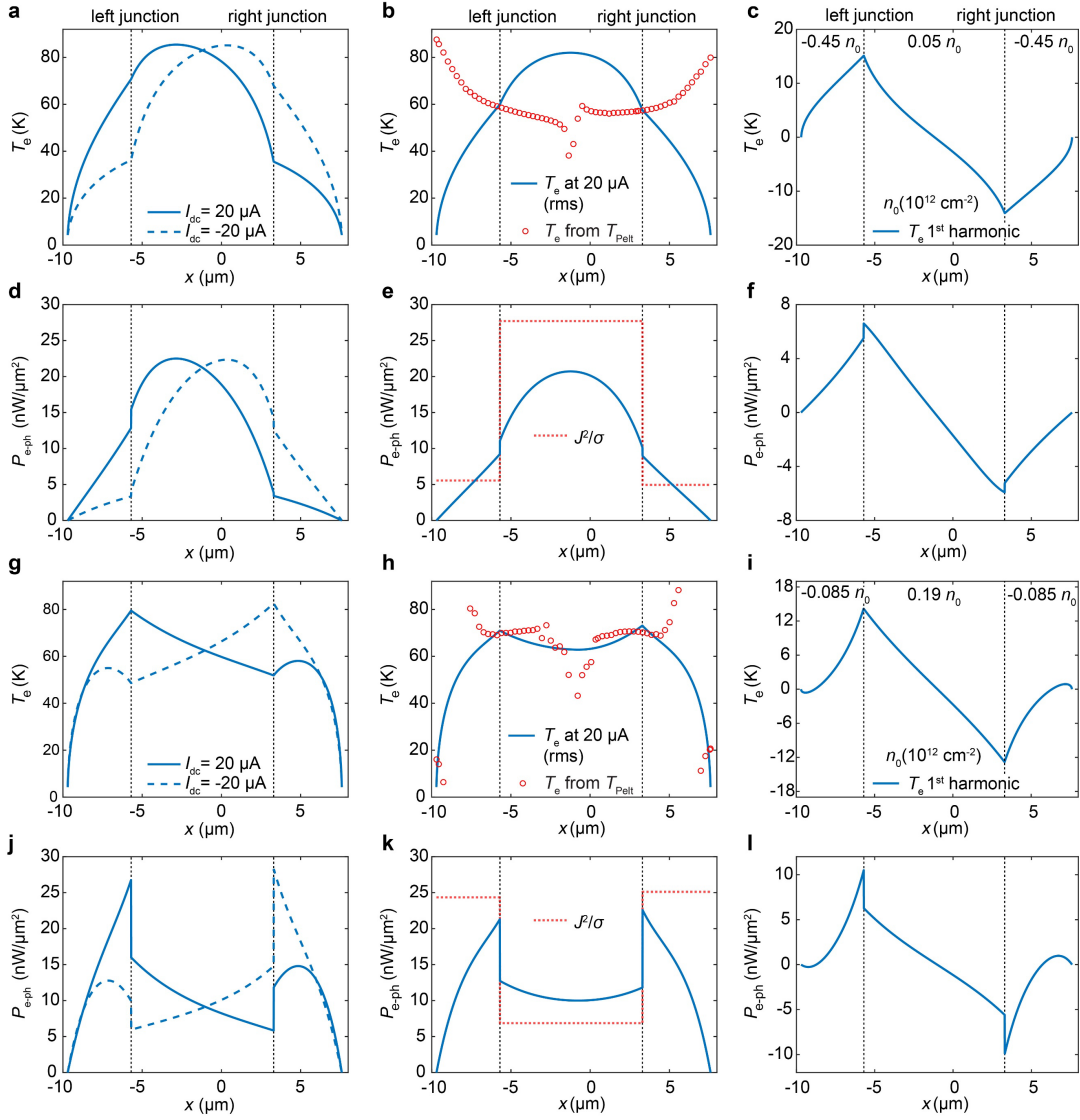

**Extended Data Fig. 6. Numerical simulations of electron temperature in p-n-p doped graphene strip.**
**a,** Numerically simulated electron temperature profile $T_e(x)$ (blue curves) for p-n-p doping of $(-0.45, 0.05, -0.45) \times 10^{12}$ cm$^{-2}$ and $\tau_i$ values of $(50, 12, 50)$ ps for *dc* current $I_{dc} = 20$ μA (solid curve) and $I_{dc} = -20$ μA (dashed curve). **b,** Profile of the rms value of $T_e(x)$ for *ac* current of 20 μA rms (blue curve) for the same parameters as in (a) used to generate Extended Data Figs. 3b,d. The open red circles show $T_e$ derived from COMSOL calculated surface lattice temperature $T_{\text{Pelt}}$ using Eq. 15, which coincides with the actual local $T_e$ at the junctions. **c,** The first harmonic of $T_e$ for the same parameters as in (b). **d-f,** Numerically calculated electron-phonon heat transfer rate $P_{e-ph}(x)$ (blue curves) for the same condition as in a-c. The red dotted line in (e) shows the Joule heat density $J^2/\sigma$ generated by the applied current. **g-l,** Same as (a-i) for p-n-p doping of $(-0.085, 0.19, -0.085) \times 10^{12}$ cm$^{-2}$ and $\tau_i$ of $(14, 35, 12)$ ps used for generating Figs. 2e,h.

Next, we use the Peltier thermometry method to derive $T_e$ independently from the calculated lattice temperature $T_{\text{Pelt}}$ using Eq. 15. Although this derivation is strictly valid and meaningful only at the junctions, it is informative to inspect the $T_e(x)$ calculated from $T_{\text{Pelt}}(x)$ along the entire graphene strip as shown by the red circles in Extended Data Fig. 6b. Strikingly, the red symbols coincide with the blue line at the two junctions and diverge elsewhere, providing additional confirmation of the validity of the method. The slight right/left asymmetry in the data arises from the small asymmetry in the device parameters.

Although $T_e(x)$ is continuous across the junctions, counterintuitively, the corresponding phonon heat transfer rate $P_{e-ph}(x)$ to the substrate is not, as shown in Extended Data Figs. 6d,e. This is because the collision integral in Eq. M6 depends on graphene DOS and $\tau_i$, which are discontinuous across the junction due to the difference in doping. The jumps at the junctions in Extended Data Fig. 6e show that the central graphene region has a larger $P_{e-ph}$. Since the central region has a lower carrier density and hence lower DOS, one would expect a lower $P_{e-ph}$ there based on Eq. M6. However, as shown below, the lower doping gives rise to much shorter $\tau_i$ resulting in higher $P_{e-ph}$ in the central region. This discontinuity in the electron-phonon coupling is also observed in the first-harmonic heat transfer rate to the substrate as seen in Extended Data Fig. 6f.

It is interesting to note that only a fraction of the Joule heat $J^2/\sigma$ deposited in the sample (red dotted line in Extended Data Fig. 6e) is directly transferred to phonons via $P_{e-ph}$, while the remaining heat is carried laterally by electrons towards the Au contacts. In the central region $J^2/\sigma > P_{e-ph}$, but in the outer regions near the junctions, paradoxically, $P_{e-ph} > J^2/\sigma$, meaning more heat is transferred to phonons than is locally generated through Joule heating. This excess heat originates from Joule energy dissipated in the central region, which is transported laterally by electrons and subsequently released to phonons in the outer regions.

A contrasting case is observed in a p–n–p configuration with low doping in the outer regions and high doping in the center, as shown in Extended Data Figs. 6g–l. This configuration features carrier densities $(-0.085, 0.19, -0.085) \times 10^{12}$ cm$^{-2}$ and $\tau_i$ of $(14, 35, 12)$ ps, matching the parameters used in Fig. 2. In this scenario, most of the Joule heating $J^2/\sigma$ occurs in the outer regions where $J^2/\sigma > P_{e-ph}$. In contrast, the central region experiences relatively low local Joule heating, yet $P_{e-ph} > J^2/\sigma$ due to heat conducted inward by electrons from the outer regions (Extended Data Fig. 6k). Consequently, both $T_e$ and $P_{e-ph}$ exhibit pronounced peaks at the junctions. These peaks translate into corresponding maxima in the lattice temperature $T_{\text{Joule}}$ as shown in Figs. 2g,h. Importantly, these junction peaks are not due to localized Joule heating at the junctions themselves; rather, they result from the convergence of lateral heat flow toward the center and an abrupt reduction in electron-phonon coupling (i.e., a larger $\tau_i$) in the highly doped central region.

In Extended Data Fig. 7, we plot the numerically simulated $T_e(x)$ profiles for the doping configurations presented in Figs. 4h,j with highly p-doped outer regions with $n = -0.43 \times 10^{12}$ cm$^{-2}$. The top row shows p-n-p configuration with various $n_c$ doping for *dc* current of $\pm 20$ μA, while the second row presents $T_e(x)$ for *ac* current of 20 μA rms. The red circles present the $T_e$ derived indirectly from the simulated lattice temperature $T_{\text{Pelt}}$ using Eq. 15, showing an excellent match at the junctions. Similar results for p-p'-p configurations with comparable doping levels are presented in the two bottom rows. With increased doping, the electron temperature drops due to enhanced conductivity and correspondingly reduced Joule heating. In addition, the Peltier effect drops with doping due to reduced $\Pi$ as seen by the difference in $T_e$ between positive and negative *dc* currents.

To further demonstrate the validity of the electron temperature extraction procedure, we simulated the case of a single p-n junction, as shown in Extended Data Fig. 8 for various dopings. In this configuration, again, an excellent match between the direct and indirect calculation of $T_e$ is obtained at the p-n junction.

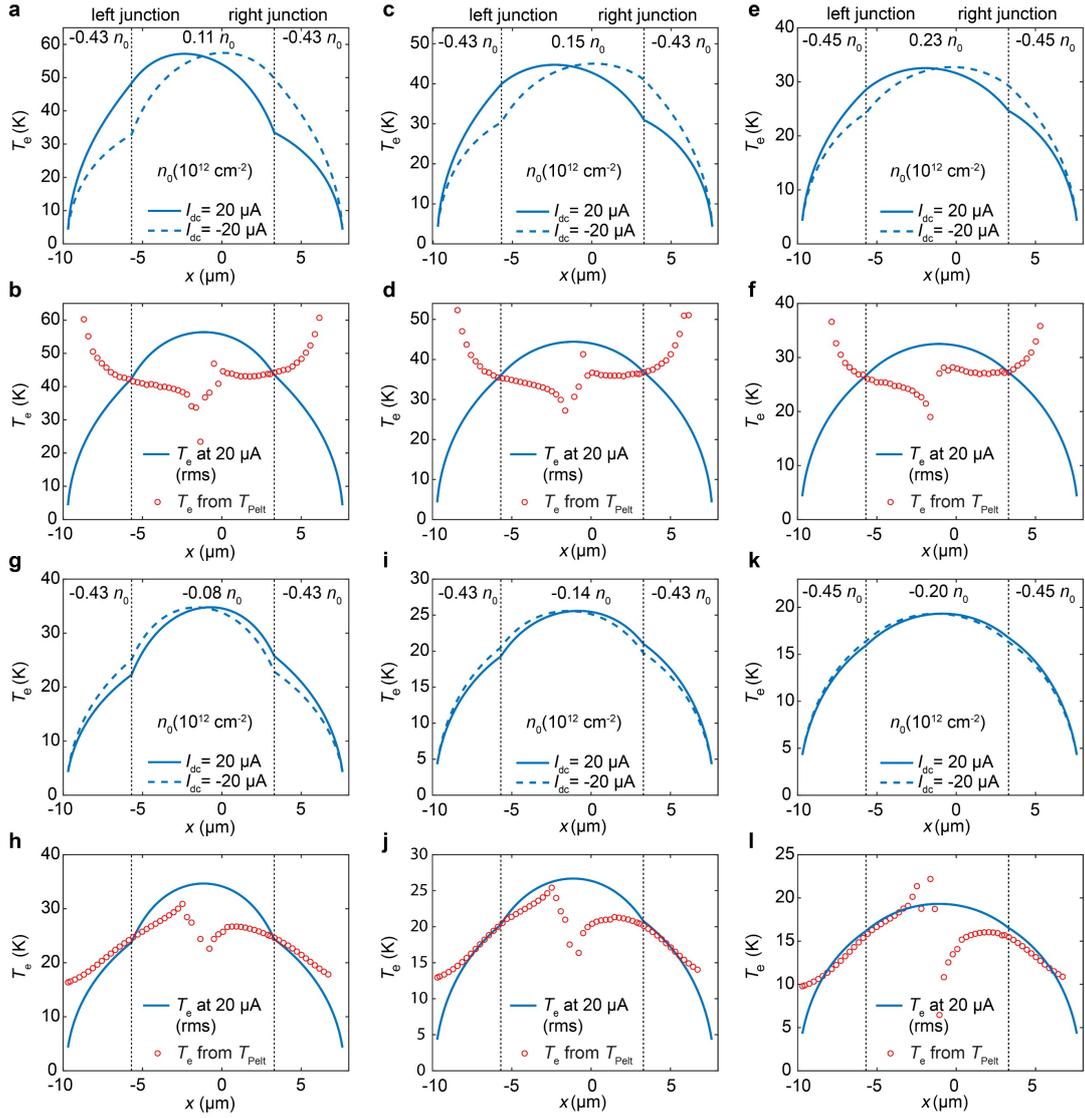

**Extended Data Fig. 7. Numerical simulation of electron temperature for additional p-n-p and p-p'-p doping configuration. a,** Numerically simulated $T_e(x)$ for p-n-p doping of $(-0.43, 0.11, -0.43) \times 10^{12}$ cm$^{-2}$ for *dc* current $I_{dc} = 20$ µA (solid curve) and $I_{dc} = -20$ µA (dashed curve). **b,** Profile of the rms value of $T_e(x)$ for *ac* current of 20 µA rms (blue curve) for the same parameters as in (a). The red circles show $T_e$ derived from COMSOL simulated lattice temperature $T_{\text{Pelt}}$. **c,d,** Same as (a,b) for the p-n-p doping $(-0.43, 0.15, -0.43) \times 10^{12}$ cm$^{-2}$. **e,f,** Same as (a,b) for $(-0.43, 0.23, -0.43) \times 10^{12}$ cm$^{-2}$. **g,h,** Same as (a,b) for p-p'-p doping $(-0.43, -0.08, -0.43) \times 10^{12}$ cm$^{-2}$. **i,j,** Same as (g,h) for $(-0.43, -0.14, -0.43) \times 10^{12}$ cm$^{-2}$. **k,l,** Same as (g,h) for $(-0.43, -0.20, -0.43) \times 10^{12}$ cm$^{-2}$.

**Derivation of material parameters**

To simulate our experimental data, we used the following materials parameters. The total bottom hBN thicknesses in the central and right/left graphene regions were 110 and 58 nm, respectively. We are unaware of experimental evaluation of hBN thermal conductivity at low temperatures. By extrapolating the published temperature dependence of bulk hBN thermal conductivity [68–70] to 4 K, the in-plane and out-of-plane thermal conductivities were taken to be $\kappa_{in-hBN} = 2$ Wm$^{-1}$K$^{-1}$ and $\kappa_{out-hBN} = 0.0075$ Wm$^{-1}$K$^{-1}$. The thermal conductivity of 70 nm thick Au metal leads was taken as $\kappa_{Au} = 10$ Wm$^{-1}$K$^{-1}$ [71,72]. The thermal conductivity of He exchange gas surrounding the 3D device was assumed to be $\kappa_{He} = 0.004$ Wm$^{-1}$K$^{-1}$ [31]. The main fitting parameter in the simulations was the doping-dependent inelastic scattering time $\tau_i$ in graphene with the best fit values presented in Fig. 4n.

The extracted $\tau_i$ is in the same range as the energy relaxation time reported in graphene for excitation energy below the optical phonon energy as studied using optical pump spectroscopy [73] and photo-thermoelectric experiments [74].

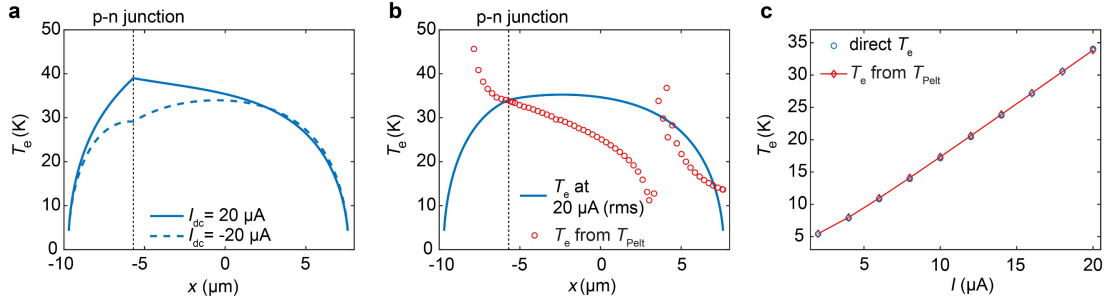

**Extended Data Fig. 8. Numerical simulation of electron temperature for a single p-n junction. a,** Numerically simulated $T_e(x)$ for p-n doping of $(-0.20, 0.20) \times 10^{12}$ cm$^{-2}$ for *dc* current $I_{dc} = 20$ µA (solid curve) and $I_{dc} = -20$ µA (dashed curve). **b,** Profile of the rms value of $T_e(x)$ for *ac* current of 20 µA rms (blue curve) for the same parameters as in (a). The red circles show $T_e$ derived from COMSOL simulated surface lattice temperature $T_{\text{Pelt}}$. **c,** The rms $T_e$ at the graphene p-n junction vs. the rms current $I$. The blue circles show $T_e$ derived directly from the electronic simulation, and the red diamonds show the $T_e$ derived indirectly from $T_{\text{Pelt}}$.

**Corrections to electron and Peltier temperatures in presence of *ac* current**

Equation 15 is strictly correct only for a *dc* current, but experimentally $T_{\text{Pelt}}(I)$ cannot be readily separated from $T_{\text{Joule}}(I)$ in a *dc* measurement. We therefore use an *ac* current at frequency $\omega = 2\pi f$, $I(t) = I_0 \sin \omega t$. In this case some numerical corrections to the derived $T_{\text{Pelt}}$ and $T_e$ are required due to harmonic expansion. The time dependent *ac* current is given by $I(t) = I_1 \sin \omega t = \sqrt{2} I \sin \omega t$, with $I_1$ and $I$ being peak and rms values, respectively. We then describe the time dependent electron temperature using Taylor expansion,

$$T_e^2(I(t)) = T_0^2 + f(I(t)) = T_0^2 + a_1 I(t) + a_2 I(t)^2 + a_3 I(t)^3. \tag{M16}$$

Since Joule heating is quadratic in $I(t)$ and Peltier shows linear plus cubic behavior, we keep here only the lowest three expansion terms for simplicity. Consequently, the time dependent $T_{\text{Pelt}}(t) = C\tilde{\Pi}_j T_e^2(I(t))I(t)$, will be described by

$$\frac{T_{\text{Pelt}}(t)}{C\pi_j} = T_e^2(I(t))I(t) = T_0^2 I + a_1 I^2 + a_2 I^3 + a_3 I^4 =$$

$$= T_0^2 I_1 \sin \omega t + a_1 I_1^2 \sin^2 \omega t + a_2 I_1^3 \sin^3 \omega t + a_3 I_1^4 \sin^4 \omega t =$$

$$= T_0^2 I_1 \sin \omega t + \frac{1}{2} a_1 I_1^2 (1 - \cos 2\omega t) + \frac{1}{4} a_2 I_1^3 (3 \sin \omega t - \sin 3\omega t) +$$

$$+ \frac{1}{8} a_3 I_1^4 (3 - 4 \cos 2\omega t + \cos 4\omega t) =$$

$$= \frac{1}{2} a_1 I_1^2 + \frac{3}{8} a_3 I_1^4 + \left(T_0^2 I_1 + \frac{3}{4} a_2 I_1^3\right) \sin \omega t - \left(\frac{1}{2} a_1 I_1^2 + \frac{1}{2} a_3 I_1^4\right) \cos 2\omega t + \cdots \tag{M17}$$

In our experimental scheme, $T_{\text{Pelt}}(I)$ is the rms value of the first harmonic $\sin \omega t$ of the measured local lattice temperature, namely

$$T_{\text{Pelt}}(I) = C\tilde{\Pi}_j \frac{\sqrt{2}}{2} \left(T_0^2 I_1 + \frac{3}{4} a_2 I_1^3\right) = C\tilde{\Pi}_j T_0^2 \left(I + \frac{3a_2}{2T_0^2} I^3\right), \tag{M18}$$

where $I = I_1/\sqrt{2}$ is the rms value of the current.

Our experimental goal is to derive the rms value $T_e$ of the electron temperature, that is the square root of the time-averaged $T_e^2(I(t))$, which in the case of sinusoidal current reduces to

$$T_e^2(I) = T_0^2 + a_2 I^2 \tag{M19}$$

from Eq. M16. Combining Eqs. M18 and M19, we attain

$$T_e = T_0\sqrt{1 + \frac{a_2}{T_0^2}I^2} = T_0\sqrt{1 + \frac{2}{3}\left(\frac{T_{\text{Pelt}}(I)/I}{\partial T_{\text{Pelt}}(I)/\partial I|_{I=0}} - 1\right)} = T_0\sqrt{1 + \frac{2b}{3}I^2}, \tag{M20}$$

which shows a factor of 2/3 correction relative to the *dc* case in Eqs. 15 and 16 due to the harmonic expansion. All the presented data analyses use Eq. M20 for derivation of the rms temperature $T_e$.